\documentclass[12pt, journal, draftclsnofoot, onecolumn]{IEEEtran}

\newtheorem{theorem}{Theorem}
\newtheorem{proposition}{Proposition}
\newtheorem{lemma}{Lemma}

\usepackage{booktabs}
\usepackage{bm}
\usepackage{graphicx}
\usepackage[cmex10]{amsmath}
\usepackage{cite}
\usepackage{amssymb}
\usepackage{subfigure}
\usepackage{extarrows}
\usepackage[letterspace = -2]{microtype}
\usepackage{algorithm}
\usepackage{algpseudocode}
\usepackage{algpascal}
\usepackage{float}
\usepackage{color, soul}

\newfloat{algorithm}{t}{lop}

\begin{document}

\title{
Relaying Strategies for Wireless-Powered MIMO Relay Networks
}

\author{
Yang~Huang,~\IEEEmembership{Student Member,~IEEE},
and Bruno~Clerckx,~\IEEEmembership{Member,~IEEE}

\thanks{A preliminary version of this paper has appeared in the IEEE International Conference on Communications 2015\cite{HC15_ICC}.}

\thanks{Y. Huang and B. Clerckx are with the Department of Electrical and Electronic Engineering, Imperial College London, London SW7 2AZ, United Kingdom (e-mail: \{y.huang13, b.clerckx\}@imperial.ac.uk). B. Clerckx is also with the School of Electrical Engineering, Korea University, Korea. This work has been partially supported by the EPSRC of UK, under grant EP/M008193/1. The work of Y. Huang was supported by China Scholarship Council (CSC) Imperial Scholarship.}

}

\maketitle

\begin{abstract}
This paper investigates relaying schemes in an amplify-and-forward multiple-input multiple-output relay network, where an energy-constrained relay harvests wireless power from the source information flow and can be further aided by an energy flow (EF) in the form of a wireless power transfer at the destination. However, the joint optimization of the relay matrix and the source precoder for the energy-flow-assisted (EFA) and the non-EFA (NEFA) schemes is intractable. The original rate maximization problem is transformed into an equivalent weighted mean square error minimization problem and optimized iteratively, where the global optimum of the nonconvex source precoder subproblem is achieved by semidefinite relaxation and rank reduction. The iterative algorithm finally converges. Then, the simplified EFA and NEFA schemes are proposed based on channel diagonalization, such that the matrices optimizations can be simplified to power optimizations. Closed-form solutions can be achieved. Simulation results reveal that the EFA schemes can outperform the NEFA schemes. Additionally, deploying more antennas at the relay increases the dimension of the signal space at the relay. Exploiting the additional dimension, the EF leakage in the information detecting block can be nearly separated from the information signal, such that the EF leakage can be amplified with a small coefficient.
\end{abstract}


\section{Introduction}
Sensor networks have been widely applied to structural monitoring, habitat monitoring, etc. Sensors may be deployed to inaccessible places, which makes replacing the sensor batteries inconvenient. In such networks, the energy of the nodes frequently selected as relays drains more quickly. The lifetime of such energy-constrained relays becomes the bottleneck to prolong the lifetime of the whole network. As a recent solution, the nodes able to harvest energy from the ambient environment are employed as relays\cite{SK11}. Nevertheless, the relay may harvest power from a more reliable and controllable energy source in the uplink transmission scenario where the destination is a collect and process center (which has a sustainable power supply). Motivated by this scenario, the paper investigates the simultaneous wireless information and power transfer (SWIPT)\cite{ZH13} in an autonomous one-way relay network, where the autonomous relay can extract energy from the incoming signal from the source to forward information but also can be aided by a dedicated power transfer from the destination. Note that the power required for CSIT sharing, etc. is not supplied by the harvested power \cite{ZPZL15}, and may come from an independent battery.

State-of-the-art SWIPT techniques for relaying can be mainly categorized into the power splitting (PS) relaying and the time switching (TS) relaying \cite{NZDK13, JZ14, DPEP13, NNZKD14arXiv, CLJMV15, ZPZL15}. Ref. \cite{NZDK13} proposed a PS relaying (where the relay extracts power for forwarding from the source information signal) and a TS relaying (where the relay harvests power from an energy signal sent by the source and then relays source information in a time-division manner). Another TS relaying, where the energy signal is sent by the destination, was studied in\cite{JZ14}. The PS relaying was also studied in the multi-pair one-way relay networks \cite{DPEP13} and the relay interference channels\cite{NNZKD14arXiv, CLJMV15}. In \cite{ZPZL15}, the relay employs dedicated antennas to harvest wireless power, while the other antennas perform PS to relay a single data stream. In the above works, the PS relaying reduces the information power at the relay. The TS relaying consumes more timeslots, though the wireless power is harvested in a dedicated timeslot. Therefore, these two methods may degrade the rate performance, and a relaying strategy able to harvest sufficient forwarding power without consuming more timeslots would be appealing.

\begin{figure}[!t]
\centering
\subfigure[Energy-flow-assisted two-phase relaying.]{
\includegraphics[width = 2.6in]
{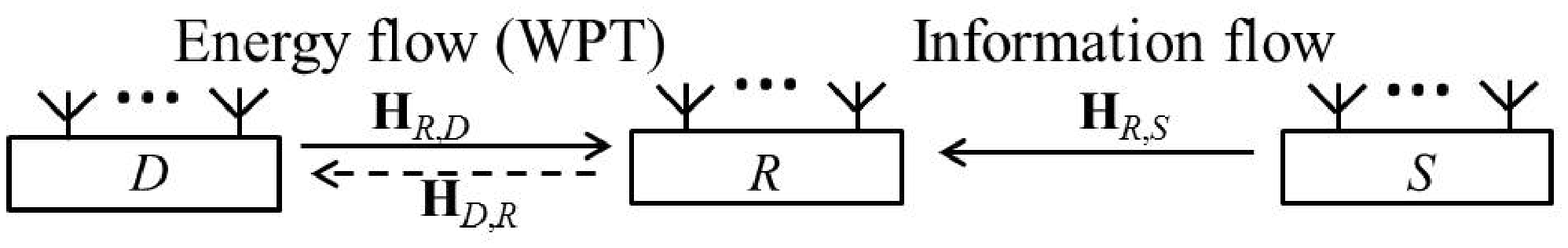}
}
\hfil
\subfigure[Two-phase relaying without energy flow.]{
\includegraphics[width = 2.9in]
{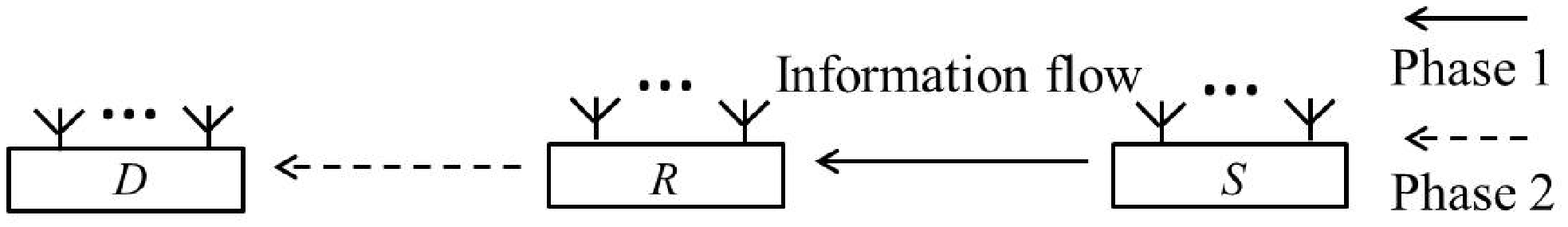}
}
\caption{SWIPT relay network. The source, relay, and destination are designated as $S$, $R$, and $D$, respectively.}
\label{FigRelayingSchemes}
\end{figure}

To circumvent those limitations, an energy-flow-assisted (EFA) two-phase amplify-and-forward (AF) one-way relaying can be proposed, where the EFA relay can harvest power from both the source information signal and a dedicated energy flow (in the form of a wireless power transfer) at the destination, as shown in Fig. \ref{FigRelayingSchemes}(a). Thanks to the PS scheme\cite{ZH13}, the received superposed signal at $R$ in phase 1 is split for information detecting (ID) and energy harvesting (EH). The energy flow (EF) leaking into the ID receiver is referred to as the EF leakage. Our previous work \cite{HC15_CommLett} shows that EF is beneficial to the EFA relay (with single antenna terminals) only in the presence of multiple relay antennas. Unfortunately, the method proposed in \cite{HC15_CommLett} cannot be exploited in the MIMO case. Here we study a more general scenario where the terminals are equipped with $r$ antennas (where $r$ is no greater than the number of relay antennas $r_R$ \cite{WT12}). The $r$-antenna terminals can transmit multiple data streams and increase the energy harvested at R through beamforming. As the harvested power at $R$ is also consumed to amplify and forward the EF leakage, the information forwarding power would be reduced, which may degrade the rate. Thus, we also investigate the non-EFA (NEFA) relaying, which harvests power from the information signal, as shown in Fig. \ref{FigRelayingSchemes}(b). The autonomous relay makes the EFA and the NEFA relaying different from the conventional relaying (where $R$ has constant energy source). The latter allocates the source power to all the diagonalized channels, only to maximize the rate\cite{RTH09}. However, if only aimed at increasing the forwarding power at $R$, NEFA would perform rank-1 beamforming at the source\cite{PC13}. That is, enhancing information transfer may conflict with enhancing energy harvesting. For EFA, the relay matrix also has to address the superposed EF leakage in the ID receiver. The main contributions of this paper are listed as follows.

Firstly, this paper proposes the EFA scheme for the multiple-input multiple-output (MIMO) relay network, where the autonomous relay is able to harvest EF from the destination and simultaneously receive the information signal from the source.

Secondly, an iterative optimization algorithm is proposed for both the EFA and NEFA schemes to jointly optimize the relay processing matrix and the source precoder. The original problem is nonconvex and intractable, which is then transformed into an equivalent problem \cite{SRLH11}, such that the matrices can be optimized iteratively. The subproblem of source precoding is essentially a nonconvex quadratically constrained quadratic problem (QCQP). As a prevailing solution, the successive convex approximation \cite{MW78, MHGKS15} cannot guarantee this subproblem yielding a global optimum, which may make the overall iterative algorithm fail to converge. To solve this problem, we formulate the nonconvex QCQP as a semidefinite program (SDP) by performing semidefinite relaxation (i.e. relaxing the nonconvex rank-1 constraint)\cite{LMSYZ10}. We show that there exists a rank-1 solution and the relaxation is safe. The global optimum (i.e. the rank-1 solution) of the original nonconvex QCQP can be derived from the solution to the SDP by performing post-processing. Finally, the iterative algorithm is shown to converge.

Thirdly, although the weighted mean-square error (WMSE) criterion and the alternating optimization (AO) have been exploited in the joint optimization of the relay networks, the issue of convergence has not been well studied. For instance, containing subproblems with multiple solutions, \cite{TSH13} only conjectures that the algorithms converge to stationary points. In this paper, supposing that tie-breaking strategies\cite{BH03} are included in solving the subproblems with multiple solutions, we prove that the minimizers converge to a limit point. This limit point is not necessarily a stationary point.

Fourthly, aiming at less complex EFA relaying algorithms, simplified algorithms are proposed. The original matrices optimization is simplified to a power optimization by performing a channel diagonalization based on the harvested-power-maximization power-leakage-minimization (HPM-PLM) strategy. Power allocation at $R$ and $S$ are optimized based on an AO. Channel pairing issues introduced in the relay power optimization are solved by an ordering operation. Closed-form solutions can be achieved in the subproblems of relay optimization and source optimization. A simplified NEFA relaying algorithm is also investigated. Simulation results show that the EF is beneficial to the EFA schemes, such that EFA schemes can outperform rate-wise NEFA schemes. Although the data streams to be forwarded are corrupted by the EF at $R$ in the EFA scheme, the antenna configuration $r_R > r$ can make the EF leakage nearly separated from the linearly combined data streams. Thus, the desired signals can be amplified with a larger coefficient.

The remainder of this paper is organized as follows. Section \ref{SecSystemModel} formulates the system model. Section \ref{SecIterAlgo} proposes the iterative algorithm for the EFA and NEFA schemes. Section \ref{SecHPM-PLM_Relaying} studies the simplified EFA schemes. Then, the simplified NEFA scheme is investigated in Section \ref{SecSimpRelayingWoEF}. Section \ref{SecSimResults} discusses the simulation results. Finally, conclusions are drawn in Section \ref{SecConclusion}.

Notations: Matrices and vectors are in bold capital and bold lower cases, respectively. The notations $(\cdot)^T$, $(\cdot)^\star$, $(\cdot)^\ast$, $(\cdot)^H$, $\text{Tr}\{\cdot\}$, $\det(\cdot)$, $\lambda_i(\cdot)$ and $[\cdot]_i$ represent the transpose, optimal solution, conjugate, conjugate transpose, trace, determinant, the $i$\,th eigenvalue and the $i$\,th column of a matrix, respectively. The notation $\mathbf{A} \succeq 0$ means that $\mathbf{A}$ is positive-semidefinite; $\pi(\mathbf{a})$ denotes the permutation; $\|\mathbf{a}\|$ denotes the 2-norm. When $\gtrless$ and $\lessgtr$ are used, top cases or bottom cases in the two notations hold simultaneously. The notation $\otimes$ denotes the Kronecker product.

\section{System Model and Problem Formulation}
\label{SecSystemModel}
As shown in Fig. \ref{FigRelayingSchemes}(a), it is considered that there is no direct link between $S$ and $D$ due to barriers (which causes huge shadow fading), such that the communication between $S$ and $D$ has to rely on the autonomous relay $R$. The $D$-to-$R$, $S$-to-$R$, and $R$-to-$D$ channels are respectively designated as {\small$\mathbf{H}_{R,D} \in \mathbb{C}^{r_R \times r}$}, {\small$\mathbf{H}_{R,S} \in \mathbb{C}^{r_R \times r}$}, and {\small$\mathbf{H}_{D,R} \in \mathbb{C}^{r \times r_R}$}, which are independent and identically distributed Rayleigh flat fading channels. Channel reciprocity is assumed such that {\small$\mathbf{H}_{D,R} = \mathbf{H}_{R,D}^T$}. It is assumed that each node has perfect full CSIT, following similar systems \cite{ZPZL15, WT12}. At each antenna of the relay, a fraction of the received power, denoted as the PS ratio $\rho$, is conveyed to the EH receiver. The noise at the ID receiver (at $R$) and $D$ are respectively denoted by {\small$\mathbf{n}_R \sim \mathcal{CN}(0, \sigma_n^2\mathbf{I})$} and {\small$\mathbf{n}_D \sim \mathcal{CN}(0, \sigma_n^2\mathbf{I})$}, while the effect of noise at the EH receiver is small and neglected \cite{ZH13}. The relay node works in a half-duplex mode.

In phase 1, the received signal at the EH receiver is given by {\small$\mathbf{y}_{R,\text{EH}} = \rho^{1/2}(\mathbf{H}_{R, D} \mathbf{B}_D \mathbf{x}_D + \mathbf{H}_{R, S} \mathbf{B}_S \mathbf{x}_S)$}, where $\mathbf{B}_D$ and $\mathbf{B}_S$ represent precoders at $D$ and $S$, respectively; $\mathbf{x}_D$ and $\mathbf{x}_S$ are transmitted signals from $D$ and $S$, respectively. Assuming an RF-to-DC conversion efficiency of 1, the harvested power equals {\small$\text{Tr}\left\{\rho\mathbf{H}_{R, D} \mathbf{Q}_D \mathbf{H}_{R, D}^H + \rho\mathbf{H}_{R, S} \mathbf{Q}_S \mathbf{H}_{R, S}^H\right\}$}, where {\small$\mathbf{Q}_D = \mathbf{B}_D \mathbf{B}_D^H$}, {\small$\mathbf{Q}_S = \mathbf{B}_S \mathbf{B}_S^H$}, {\small$\text{Tr}\{\mathbf{Q}_D\} \leq P_D$}, and {\small$\text{Tr}\{\mathbf{Q}_S\} \leq P_S$}. Meanwhile, the baseband signal input to the ID receiver for forwarding is given by {\small$\mathbf{y}_{R,\text{ID}} = (1 - \rho)^{1/2}(\mathbf{H}_{R, D} \mathbf{B}_D \mathbf{x}_D + \mathbf{H}_{R, S} \mathbf{B}_S \mathbf{x}_S) + \mathbf{n}_R$}. In phase 2, the information received at $D$ is given by {\small$\mathbf{y}_D = (1 - \rho)^{1/2}\mathbf{H}_{D, R}\mathbf{F} (\mathbf{H}_{R, S} \mathbf{B}_S \mathbf{x}_S + \mathbf{H}_{R, D} \mathbf{B}_D \mathbf{x}_D ) + \mathbf{H}_{D, R}\mathbf{F}\mathbf{n}_R + \mathbf{n}_D$}, where $\mathbf{F}$ denotes the relay processing matrix. With perfect CSI at $D$, following the similar systems\cite{WT12}, we assume that the self-interference in $\mathbf{y}_D$, i.e. the term related to $\mathbf{x}_D$, can be canceled, but power at the relay is consumed to forward this self-interference (i.e. the EF leakage). Defining {\small$\mathbf{R} = \sigma_{n}^2\mathbf{H}_{D, R} \mathbf{F} \mathbf{F}^H \mathbf{H}_{D, R}^H + \sigma_{n}^2\mathbf{I}$}, the rate maximization problem can be formulated as
{\small
\begin{IEEEeqnarray}{cl}
\label{ProblemP1}
\max_{\mathbf{B}_S, \mathbf{F}}\, & \frac{1}{2}\log\det \left(\mathbf{I} \! + \! (1 \! - \! \rho)\mathbf{H}_{D, R} \mathbf{F} \mathbf{H}_{R, S} \mathbf{B}_S \mathbf{B}_S^H \mathbf{H}_{R, S}^H \mathbf{F}^H \mathbf{H}_{D, R}^H \mathbf{R}^{-1} \right) \IEEEyessubnumber \label{C_NonIF_P1}\\
\text{s.t.} & \text{Tr}\left\{ (1 \! - \! \rho)\left(\mathbf{F}\mathbf{H}_{R, S} \mathbf{B}_S \mathbf{B}_S^H \mathbf{H}_{R, S}^H \mathbf{F}^H \! + \! \mathbf{F} \mathbf{H}_{R, D} \mathbf{Q}_D \mathbf{H}_{R, D}^H \mathbf{F}^H\right) + \sigma_n^2\mathbf{F}\mathbf{F}^H \! \right\} \! \leq \nonumber\\
& \rho \text{Tr}\left\{ \! \mathbf{H}_{R, D} \mathbf{Q}_D \mathbf{H}_{R, D}^H \! + \! \mathbf{H}_{R, S} \mathbf{B}_S \mathbf{B}_S^H \mathbf{H}_{R, S}^H \! \right\},\IEEEyessubnumber \label{FwdPowerConst_1P1} \\
& \text{Tr}\{\mathbf{B}_S\mathbf{B}_S^H\} \leq P_S\,.\IEEEyessubnumber \label{SrcPowerConst_1P1}
\end{IEEEeqnarray}
}In this problem, the optimization is not performed over $\mathbf{Q}_D$, due to the high difficulty\footnote{Since $\mathbf{Q}_D$ is absent from (\ref{C_NonIF_P1}), it cannot be optimized iteratively. Alternatively, if the coupled $\mathbf{F}$ and $\mathbf{Q}_D$ are optimized in a subproblem, the subproblem is essentially a bilinear problem, which is NP-hard and hard to yield the global optimum\cite{VBS99}. This means that the value of (\ref{C_NonIF_P1}) may not monotonically increases over iterations, and the iterative algorithm cannot converge.}. In the following sections, the iterative algorithm for both the EFA and NEFA schemes is proposed, where the corresponding algorithms are designated as EFA-OPT and NEFA-OPT. The algorithm iteratively optimizes $\mathbf{B}_S$ and $\mathbf{F}$. Then, to avoid the complexity of matrices optimization, a simplified EFA algorithm (designated as EFA-S1) is proposed based on channel diagonalization, such that the problem is simplified to a power optimization. This algorithm is further simplified as EFA-S2 and provide closed-form solutions of the relay and source strategies. Finally, a simplified NEFA algorithm (designated as NEFA-S) is proposed by using channel diagonalization.

\section{Iterative Algorithm}
\label{SecIterAlgo}
This section proposes an iterative algorithm to solve the joint optimization problem (\ref{ProblemP1}), where the value of the fixed $\mathbf{Q}_D$ depends on the relaying scheme and differs in EFA-OPT and NEFA-OPT. Take the singular value decomposition (SVD) of $\mathbf{H}_{R,D}\!=\!\mathbf{V}_{D,R}^{\ast} \mathbf{\Sigma}_{D,R} \mathbf{U}_{D,R}^T$ (due to the channel reciprocity). In EFA-OPT, to maximize the amount of power harvested at $R$, $\mathbf{Q}_D \! =\! P_D [\mathbf{U}_{D,R}^{\ast}]_{\text{max}} [\mathbf{U}_{D,R}^{\ast}]_{\text{max}}^H$, where $[\mathbf{U}_{D,R}^{\ast}]_{\text{max}}$ is the right singular vector (RSV) corresponding to the maximum singular value $\lambda_{D,R,\text{max}}^{1/2}$ of $\mathbf{H}_{R,D}$ (see Proposition 1 in \cite{PC13}). In NEFA-OPT, $\mathbf{Q}_D = \mathbf{0}$.

Because the optimization variable $\mathbf{F}$ within the matrix inversion $\mathbf{R}^{-1}$ makes problem (\ref{ProblemP1}) intractable, we introduce an auxiliary variable $\mathbf{A}_0 \succeq 0$ to transform problem (\ref{ProblemP1}) into an equivalent WMSE minimization problem \cite{SRLH11} given by
{\small
\begin{IEEEeqnarray}{cl}
\label{WMSEProBsF}
\min_{\mathbf{A}_0 \succeq 0, \mathbf{W}, \mathbf{F}, \mathbf{B}_S} & \text{Tr}\left\{ \mathbf{A}_0 \mathbf{E}\left(\mathbf{W}, \mathbf{F}, \mathbf{B}_S\right) \right\} - \log \det \left(\mathbf{A}_0\right) \IEEEyessubnumber \label{EqObjWMSEProBsF}\\
\text{s.t.} & \text{Tr}\left\{ (1 - \rho)\mathbf{F}\mathbf{H}_{R, S} \mathbf{B}_S \mathbf{B}_S^H \mathbf{H}_{R, S}^H \mathbf{F}^H + (1 - \rho) \mathbf{F} \mathbf{H}_{R, D} \mathbf{Q}_D \mathbf{H}_{R, D}^H \mathbf{F}^H + \sigma_n^2\mathbf{F}\mathbf{F}^H \right\} \leq \nonumber\\
& \rho \text{Tr}\left\{\mathbf{H}_{R, D} \mathbf{Q}_D \mathbf{H}_{R, D}^H + \mathbf{H}_{R, S}\mathbf{B}_S \mathbf{B}_S^H \mathbf{H}_{R, S}^H \right\}\,, \IEEEyessubnumber \label{EqRelayPwrConstWMSEProBsF}\\
& \text{Tr}\{\mathbf{B}_S\mathbf{B}_S^H\} \leq P_S \,, \IEEEyessubnumber \label{EqSrcPwrConstWMSEProBsF}
\end{IEEEeqnarray}
}where $\mathbf{W}$ denotes the receive filter at $D$, while $\mathbf{E}(\cdot)$ represents the MSE matrix defined in the MSE {\small$\mathcal{E}\{ (\mathbf{W}^H \mathbf{y}_D - \mathbf{x}_S)^H (\mathbf{W}^H \mathbf{y}_D - \mathbf{x}_S)\} = \text{Tr}\{\mathcal{E}\{(\mathbf{W}^H \mathbf{y}_D - \mathbf{x}_S)(\mathbf{W}^H \mathbf{y}_D - \mathbf{x}_S)^H\}\} = \text{Tr}\{\mathbf{E}(\mathbf{W}, \mathbf{F}, \mathbf{B}_S)\}$}, and {\small$\mathbf{E}(\mathbf{W}, \mathbf{F}, \mathbf{B}_S)$} is given by
{\small
\begin{IEEEeqnarray}{rcl}
\mathbf{E} &{} = {}& (1 - \rho) \mathbf{W}^H \mathbf{H}_{D,R} \mathbf{F} \mathbf{H}_{R,S} \mathbf{B}_S \mathbf{B}_S^H \mathbf{H}_{R,S}^H \mathbf{F}^H \mathbf{H}_{D,R}^H \mathbf{W} + \mathbf{W}^H \mathbf{H}_{D,R} \mathbf{F} \mathbf{F}^H \mathbf{H}_{D,R}^H \mathbf{W} \sigma_n^2 \nonumber\\
&&{}+{} \mathbf{W}^H \mathbf{W} \sigma_n^2 - (1 - \rho)^{1/2} \mathbf{B}_S^H \mathbf{H}_{R,S}^H \mathbf{F}^H \mathbf{H}_{D,R}^H \mathbf{W} - (1 - \rho)^{1/2} \mathbf{W}^H \mathbf{H}_{D,R} \mathbf{F} \mathbf{H}_{R,S} \mathbf{B}_S + \mathbf{I}_r.
\end{IEEEeqnarray}
}The proof of the equivalence between problems (\ref{ProblemP1}) and (\ref{WMSEProBsF}) is similar to the Appendix A in \cite{SRLH11}. The details are omitted here. Since the optimization variables $\mathbf{A}_0$, $\mathbf{W}$, $\mathbf{F}$, and $\mathbf{B}_S$ are coupled in (\ref{EqObjWMSEProBsF}) and (\ref{EqRelayPwrConstWMSEProBsF}), problem (\ref{WMSEProBsF}) is still intractable. Subsequently, the original problem is decoupled into four subproblems of $\mathbf{A}_0$, $\mathbf{W}$, $\mathbf{F}$, and $\mathbf{B}_S$. The variable corresponding to each subproblem is alternatively optimized by fixing the others.

\subsection{Subproblems of $\mathbf{A}_0$ and $\mathbf{W}$}
Fixing the variables $(\mathbf{W}, \mathbf{F}, \mathbf{B}_S)$, the subproblems of $\mathbf{A}_0$ can be written as
{\small
\begin{equation}
\label{SubproA_0_WMSEProBsF}
\min_{\mathbf{A}_0 \succeq 0} \, \text{Tr}\{\mathbf{A}_0\mathbf{E}\} - \log\det(\mathbf{A}_0)\,.
\end{equation}
}Similarly, fixing the variables $(\mathbf{A}_0, \mathbf{F}, \mathbf{B}_S)$, the subproblem of $\mathbf{W}$ can be formulated as
{\small
\begin{equation}
\label{SubproW_WMSEProBsF}
\min_{\mathbf{W}} \, \text{Tr}\{\mathbf{A}_0 \mathbf{E}(\mathbf{\mathbf{W}})\} \,.
\end{equation}
}Since the above two subproblems are strictly convex, an unique optimal solution can be obtained for each subproblem by the first-order condition of optimality. Calculating the derivatives of objective functions of the two subproblems\cite{HG07}, the optimal $\mathbf{A}_0^\star$ and $\mathbf{W}^\star$ (which is the minimum mean square error receiver) are given by
{\small
\begin{equation}
\label{EqOptimA_0}
\mathbf{A}_0^\star = \mathbf{E}^{-1}\,,
\end{equation}
\begin{IEEEeqnarray}{rcl}
\label{EqOptimW}
\mathbf{W}^\star = \mathbf{W}_{\text{mmse}} &{}={}&  \left[(1 - \rho)\mathbf{H}_{D,R} \mathbf{F} \mathbf{H}_{R,S} \mathbf{B}_S \mathbf{B}_S^H \mathbf{H}_{R,S}^H \mathbf{F}^H \mathbf{H}_{D,R}^H + \mathbf{H}_{D,R} \mathbf{F} \mathbf{F}^H \mathbf{H}_{D,R}^H \sigma_n^2 + \mathbf{I}_r \sigma_n^2 \right]^{-1} \nonumber\\
&& {}\cdot{} (1 - \rho)^{1/2} \mathbf{H}_{D,R} \mathbf{F} \mathbf{H}_{R,S} \mathbf{B}_S \,.
\end{IEEEeqnarray}
}Substituting (\ref{EqOptimA_0}) into (\ref{EqObjWMSEProBsF}) yields {\small$\text{Tr}\{\mathbf{I}\} - \log \det (\mathbf{E}^{-1}(\mathbf{W}, \mathbf{F}, \mathbf{B}_S))$}, where {\small$\log\det(\mathbf{E}^{-1})$} is equal to twice the end-to-end achievable rate, i.e. (\ref{C_NonIF_P1}). This reveals the physical meaning of the quantity of the objective function (\ref{EqObjWMSEProBsF}) and the equivalence between problems (\ref{ProblemP1}) and (\ref{WMSEProBsF}).

\subsection{Subproblem of $\mathbf{F}$}
Fixing the variables {\small$(\mathbf{A}_0,\! \mathbf{W},\! \mathbf{B}_S\!)$}, the subproblem of $\mathbf{F}$ (where $r_R \! \geq \! r \! > \! 1$) can be formulated as
{\small
\begin{IEEEeqnarray}{cl}
\label{OrigSubproF_WMSEProBsF}
\min_{\mathbf{F}} \, & \text{Tr}\left\{ (1 - \rho)\mathbf{F}^H \mathbf{H}_{D,R}^H \mathbf{W} \mathbf{A}_0 \mathbf{W}^H \mathbf{H}_{D,R} \mathbf{F} \mathbf{H}_{R,S} \mathbf{B}_S \mathbf{B}_S^H \mathbf{H}_{R,S}^H  + \sigma_n^2 \mathbf{F}^H \mathbf{H}_{D,R}^H \mathbf{W} \mathbf{A}_0 \mathbf{W}^H \mathbf{H}_{D,R}\mathbf{F}\right. \nonumber\\
&\left.{}-{} (1 - \rho)^{1/2}\mathbf{F}^H \mathbf{H}_{D,R}^H \mathbf{W} \mathbf{A}_0 \mathbf{B}_S^H \mathbf{H}_{R,S}^H - (1 - \rho)^{1/2} \mathbf{H}_{R,S} \mathbf{B}_S \mathbf{A}_0 \mathbf{W}^H \mathbf{H}_{D,R} \mathbf{F}\right\} \\
\text{s.t.} \, & \text{(\ref{EqRelayPwrConstWMSEProBsF})} \,. \nonumber
\end{IEEEeqnarray}
}By applying the manipulation {\small$\text{Tr}\{\mathbf{A}\mathbf{B}\mathbf{C}\} = \text{vec}(\mathbf{A}^H)^H (\mathbf{I} \otimes \mathbf{B}) \text{vec}(\mathbf{C})$}, {\small$\text{vec}(\mathbf{A} \mathbf{B}) = (\mathbf{B}^T \otimes \mathbf{I}) \text{vec}(\mathbf{A})$}, and {\small$\text{Tr}\{\mathbf{A} \mathbf{B} \mathbf{C} \mathbf{D}\} = \text{vec}(\mathbf{A}^H)^H (\mathbf{D}^T \otimes \mathbf{B}) \text{vec}(\mathbf{C})$}, problem (\ref{OrigSubproF_WMSEProBsF}) can be equivalently written as
{\small
\begin{IEEEeqnarray}{cl}
\label{QCQPsubproF_WMSEProBsF}
\min_{\mathbf{f}} \, & \mathbf{f}^H \mathbf{A}_1 \mathbf{f} - \mathbf{f}^H \mathbf{a}_1 - \mathbf{a}_1^H \mathbf{f} \IEEEyessubnumber \label{EqObjQCQPsubproF_WMSEProBsF}\\
\text{s.t.} \, & \mathbf{f}^H \mathbf{A}_2 \mathbf{f} \leq C_f\,, \IEEEyessubnumber \label{EqConstQCQPsubproF_WMSEProBsF}
\end{IEEEeqnarray}
}where {\small$\mathbf{f} = \text{vec}(\mathbf{F})$}, {\small$\mathbf{A}_1 = (((1 - \rho)\mathbf{H}_{R,S} \mathbf{B}_S \mathbf{B}_S^H \mathbf{H}_{R,S}^H)^T + \mathbf{I}_{r_R}\sigma_n^2) \otimes (\mathbf{H}_{D,R}^H \mathbf{W} \mathbf{A}_0 \mathbf{W}^H \mathbf{H}_{D,R})$}, {\small$\mathbf{a}_1 = \text{vec}((1 - \rho)^{1/2} \mathbf{H}_{D,R}^H \mathbf{W} \mathbf{A}_0 \mathbf{B}_S^H \mathbf{H}_{R,S}^H)$}, {\small$\mathbf{A}_2 = ((1 - \rho)\mathbf{H}_{R,S} \mathbf{B}_S \mathbf{B}_S^H \mathbf{H}_{R,S}^H + (1 - \rho)\mathbf{H}_{R,D} \mathbf{Q}_D \mathbf{H}_{R,D}^H + \mathbf{I}_{r_R}\sigma_n^2)^T \otimes \mathbf{I}_{r_R}$}, and {\small$C_f = \rho \text{Tr}\{\mathbf{H}_{R,D} \mathbf{Q}_D \mathbf{H}_{R,D}^H + \mathbf{H}_{R,S} \mathbf{B}_S \mathbf{B}_S^H \mathbf{H}_{R,S}^H\}$}. Because of the positive-semidefinite $\mathbf{A}_1$ and $\mathbf{A}_2$, problem (\ref{QCQPsubproF_WMSEProBsF}) is a convex QCQP. Although the numerical result can be achieved by solving the problem with an convex optimization toolbox such as CVX\cite{GB14}, a closed-form solution can be obtained by analyzing the Karush-Kuhn-Tucker (KKT) conditions. Letting $\xi_1$ denote the Lagrangian multiplier associated to (\ref{EqConstQCQPsubproF_WMSEProBsF}), the KKT conditions of problem (\ref{QCQPsubproF_WMSEProBsF}) are listed as {\small$[\mathbf{f}^\star]^H \mathbf{A}_2 \mathbf{f}^\star \leq C_f$}, {\small$\xi_1^\star \geq 0$}, {\small$\xi_1^\star ([\mathbf{f}^\star]^H \mathbf{A}_2 \mathbf{f}^\star - C_f) = 0$}, and {\small$(\mathbf{A}_1 + \xi_1^\star \mathbf{A}_2)\mathbf{f}^\star = \mathbf{a}_1$}. It follows that if {\small$[\mathbf{f}^\star]^H \mathbf{A}_2 \mathbf{f}^\star < C_f$}, {\small$\xi_1^\star = 0$} and {\small$\mathbf{A}_1 \mathbf{f}^\star = \mathbf{a}_1$}; if {\small$[\mathbf{f}^\star]^H \mathbf{A}_2 \mathbf{f}^\star = C_f$}, {\small$(\mathbf{A}_1 + \xi_1^\star \mathbf{A}_2)\mathbf{f}^\star = \mathbf{a}_1$}. Thus, if $\mathbf{a}_1$ is within the column space of $\mathbf{A}_1$ and {\small$\mathbf{a}_1^H \mathbf{A}_1^\dag \mathbf{A}_2 \mathbf{A}_1^\dag \mathbf{a}_1 < C_f$} (where $\mathbf{A}_1^\dag$ denotes the pseudo inverse of $\mathbf{A}_1$), the closed-form solution is obtained as
{\small
\begin{equation}
\label{Eq_CloFormSolu_QCQPsubproF_constInactive}
\mathbf{f}^\star = \mathbf{A}_1^\dag\mathbf{a}_1 + \mathcal{N}(\mathbf{A}_1)\,,
\end{equation}
}where $\mathcal{N}(\mathbf{A}_1)$ denotes the null space of $\mathbf{A}_1$. Otherwise,
the optimal solution is given by {\small$\mathbf{f}^\star = (\mathbf{A}_1 + \xi_1^\star \mathbf{A}_2)^{-1} \mathbf{a}_1$}, where the optimal $\xi_1^\star$ can be achieved by solving $\mathbf{a}_1^H (\mathbf{A}_1 + \xi_1^\star \mathbf{A}_2)^{-1} \mathbf{A}_2 (\mathbf{A}_1 + \xi_1^\star \mathbf{A}_2)^{-1} \mathbf{a}_1 = C_f$.
In the scenario where $r_R \geq r = 1$, the rate maximization problem is equivalent to the signal-to-noise ratio (SNR) maximization, which boils down to the problem solved in \cite{HC15_CommLett}.

\subsection{Subproblem of $\mathbf{B}_S$}
Fixing the variables {\small$(\mathbf{A}_0, \mathbf{W}, \mathbf{F})$}, the subproblem of {\small$\mathbf{B}_S$} can be written as
{\small
\begin{IEEEeqnarray}{cl}
\label{OrigSubproBs_WMSEProBsF}
\min_{\mathbf{B}_S} \, & \text{Tr}\left\{ (1 - \rho)\mathbf{B}_S^H \mathbf{H}_{R,S}^H\mathbf{F}^H \mathbf{H}_{D,R}^H \mathbf{W} \mathbf{A}_0 \mathbf{W}^H \mathbf{H}_{D,R} \mathbf{F} \mathbf{H}_{R,S} \mathbf{B}_S \right\} \nonumber\\
&{}-{} \text{Tr}\left\{(1 - \rho)^{1/2}\mathbf{B}_S^H \mathbf{H}_{R,S}^H \mathbf{F}^H \mathbf{H}_{D,R}^H \mathbf{W} \mathbf{A}_0 + (1 - \rho)^{1/2} \mathbf{A}_0 \mathbf{W}^H \mathbf{H}_{D,R} \mathbf{F} \mathbf{H}_{R,S} \mathbf{B}_S\right\} \\
\text{s.t.} \, & \text{(\ref{EqRelayPwrConstWMSEProBsF}) and (\ref{EqSrcPwrConstWMSEProBsF})}\,. \nonumber
\end{IEEEeqnarray}
}Similarly to the linear algebra manipulation of problem (\ref{OrigSubproF_WMSEProBsF}), problem (\ref{OrigSubproBs_WMSEProBsF}) can be further formulated as an equivalent QCQP problem given by
{\small
\begin{IEEEeqnarray}{cl}
\label{QCQPsubproBs_WMSEProBsF}
\min_{\mathbf{b}} \, & \mathbf{b}^H \mathbf{A}_3 \mathbf{b} - \mathbf{b}^H \mathbf{a}_2 - \mathbf{a}_2^H \mathbf{b} \IEEEyessubnumber \label{EqObjQCQPsubproBs_WMSEProBsF}\\
\text{s.t.} \, & \mathbf{b}^H \mathbf{A}_4 \mathbf{b} \leq C_b\,, \IEEEyessubnumber \label{EqRelayConstQCQPsubproBs_WMSEProBsF} \\
& \mathbf{b}^H \mathbf{b} \leq P_S\,, \IEEEyessubnumber
\end{IEEEeqnarray}
}where {\small$\mathbf{b} = \text{vec}(\mathbf{B}_S)$}, {\small$\mathbf{A}_3 = \mathbf{I}_r \otimes ((1 - \rho)\mathbf{H}_{R,S}^H \mathbf{F}^H \mathbf{H}_{D,R}^H \mathbf{W} \mathbf{A}_0 \mathbf{W}^H \mathbf{H}_{D,R} \mathbf{F} \mathbf{H}_{R,S})$}, {\small$\mathbf{a}_2 = \text{vec}((1 - \rho)^{1/2} \mathbf{H}_{R,S}^H \mathbf{F}^H \cdot \mathbf{H}_{D,R}^H  \mathbf{W} \mathbf{A}_0)$}, {\small$\mathbf{A}_4 = \mathbf{I}_r \otimes ((1 - \rho)\mathbf{H}_{R,S}^H \mathbf{F}^H \mathbf{F} \mathbf{H}_{R,S} - \rho \mathbf{H}_{R,S}^H \mathbf{H}_{R,S})$}, and {\small$C_b = \text{Tr}\{\rho \mathbf{H}_{R, D} \mathbf{Q}_D \mathbf{H}_{R, D}^H - (1 - \rho) \mathbf{F} \mathbf{H}_{R, D} \mathbf{Q}_D \mathbf{H}_{R, D}^H \mathbf{F}^H - \sigma_n^2\mathbf{F}\mathbf{F}^H\}$}. Because $\mathbf{A}_4$ is indefinite, (\ref{EqRelayConstQCQPsubproBs_WMSEProBsF}) is nonconvex and problem (\ref{QCQPsubproBs_WMSEProBsF}) is nonconvex, which makes it hard to find the global optimum solution. Since the critical problem is the nonconvex (\ref{EqRelayConstQCQPsubproBs_WMSEProBsF}), in the following section, we transform problem (\ref{QCQPsubproBs_WMSEProBsF}) into an equivalent form such that (\ref{EqRelayConstQCQPsubproBs_WMSEProBsF}) can be written in a linear form and the reformulated (\ref{EqRelayConstQCQPsubproBs_WMSEProBsF}) can be convex.

\subsubsection{Convex Relaxation}
\label{SubsubsecConvxRelax}
By introducing auxiliary variables $t$ and $\mathbf{b}^\prime$ (subject to $\mathbf{b} = \mathbf{b}^\prime/t $ and $|t|^2 = 1$), problem (\ref{QCQPsubproBs_WMSEProBsF}) is transformed into an equivalently homogenized form given by
{\small
\begin{IEEEeqnarray}{cl}
\label{HQCQPsubproBs_WMSEProBsF}
\min_{\mathbf{b}^\prime, t} \, & \text{Tr}\left\{\mathbf{B}_1 \mathbf{\Phi}(\mathbf{b}^\prime, t)\right\} \IEEEyessubnumber \\
\text{s.t.} \, & \text{Tr}\left\{\mathbf{B}_2 \mathbf{\Phi}(\mathbf{b}^\prime, t) \right\} \leq C_b\,, \IEEEyessubnumber \\
& \text{Tr}\left\{\mathbf{B}_3 \mathbf{\Phi}(\mathbf{b}^\prime, t) \right\} \leq P_S\,. \IEEEyessubnumber \\
& \text{Tr}\left\{\mathbf{B}_4 \mathbf{\Phi}(\mathbf{b}^\prime, t) \right\} = 1\,, \IEEEyessubnumber
\end{IEEEeqnarray}
}where {\small$\mathbf{\Phi}(\mathbf{b}^\prime, t) = (\mathbf{b}^\prime[\mathbf{b}^\prime]^H, t^\ast \mathbf{b}^\prime; t[\mathbf{b}^\prime]^H, |t|^2)$}, {\small$\mathbf{B}_1 = (\mathbf{A}_3, - \mathbf{a}_2; - \mathbf{a}_2^H, 0)$}, {\small$\mathbf{B}_2 = (\mathbf{A}_4, \mathbf{0}; \mathbf{0}^H, 0)$}, {\small$\mathbf{B}_3 = (\mathbf{I}, \mathbf{0}; \mathbf{0}^H, 0)$}, and {\small$\mathbf{B}_4 = (\mathbf{0}_{r^2 \times r^2}, \mathbf{0}; \mathbf{0}^H, 1)$}. In problem (\ref{HQCQPsubproBs_WMSEProBsF}), the optimal $\mathbf{b}^\star$ of problem (\ref{QCQPsubproBs_WMSEProBsF}) can be achieved by calculating {\small$\mathbf{b}^\star = [\mathbf{b}^\prime]^\star /t^\star$}. In order to solve problem (\ref{HQCQPsubproBs_WMSEProBsF}), by replacing the variables $\mathbf{\Phi}(\mathbf{b}^\prime,t)$ with one matrix variable $\mathbf{X}_b$, the problem can be linearized as an equivalent form given by
{\small
\begin{IEEEeqnarray}{cl}
\label{RconstSDPsubproBs_WMSEProBsF}
\min_{\mathbf{X}_b \succeq 0} \, & \text{Tr}\left\{ \mathbf{B}_1 \mathbf{X}_b \right\} \IEEEyessubnumber\\
\text{s.t.} \, & \text{Tr}\left\{ \mathbf{B}_2 \mathbf{X}_b \right\} \leq C_b\,, \IEEEyessubnumber \label{EqRpwrConst_RconstSDP} \\
& \text{Tr}\left\{ \mathbf{B}_3 \mathbf{X}_b \right\} \leq P_S\,, \IEEEyessubnumber \label{EqSpwrConst_RconstSDP}\\
& \text{Tr}\left\{ \mathbf{B}_4 \mathbf{X}_b \right\} = 1\,,\IEEEyessubnumber \label{EqHomoConst_RconstSDP}\\
& \text{rank}\left(\mathbf{X}_b\right) = 1\,.\IEEEyessubnumber \label{EqRankConst}
\end{IEEEeqnarray}
}Note that problem (\ref{RconstSDPsubproBs_WMSEProBsF}) is still nonconvex and intractable, due to the rank constraint (\ref{EqRankConst}). In order to obtain the solution of (\ref{RconstSDPsubproBs_WMSEProBsF}), we relax (\ref{EqRankConst}), achieving a SDP given by
{\small
\begin{IEEEeqnarray}{cl}
\label{SDRsubproBs_WMSEProBsF}
\min_{\mathbf{X}_b \succeq 0} \, & \text{Tr}\left\{ \mathbf{B}_1 \mathbf{X}_b \right\} \\
\text{s.t.} \, & \text{(\ref{EqRpwrConst_RconstSDP}), (\ref{EqSpwrConst_RconstSDP}), and (\ref{EqHomoConst_RconstSDP})}. \nonumber
\end{IEEEeqnarray}
}Problem (\ref{SDRsubproBs_WMSEProBsF}) is convex and can be solved by CVX. However, the minimized value of the objective function $\text{Tr}\{ \mathbf{B}_1 \mathbf{X}_b \}$ may only provide a lower bound of the original problem, because the achieved minimizers of problem (\ref{SDRsubproBs_WMSEProBsF}) may violate the rank constraint (\ref{EqRankConst}) in the original problem. Fortunately, as proved in Proposition 3.5 in \cite{HP10}, for a separable SDP with $m_x$ matrix variables and $m_c$ linear constraints, if $m_c \leq m_x + 2$, an optimal solution to the SDP exists with each minimizer of rank one.
It can be shown that (\ref{SDRsubproBs_WMSEProBsF}) satisfies all the conditions required by Proposition 3.5 in \cite{HP10}.
Thus, problem (\ref{SDRsubproBs_WMSEProBsF}) has among others a rank-1 solution. This means that with such a rank-1 solution, (\ref{EqRankConst}) can be safely relaxed and the achieved rank-1 solution turns out to be the global optimum of problem (\ref{RconstSDPsubproBs_WMSEProBsF}). Thereby, the global optimal solutions of problems (\ref{HQCQPsubproBs_WMSEProBsF}) and (\ref{QCQPsubproBs_WMSEProBsF}) can be achieved.

\subsubsection{Postprocessing to Obtain the Rank-1 Solution}
However, it is worth noting that problem (\ref{SDRsubproBs_WMSEProBsF}) does not only have a rank-one solution, and the contemporary interior-point algorithms (IPA) (which are exploited to obtain numerical results for SDPs) usually yield highest-rank solutions\cite{Dattorro05}. That is, the optimal $\mathbf{X}_b^\star$ for (\ref{SDRsubproBs_WMSEProBsF}) achieved by CVX (or other optimization solvers based on the interior-point algorithm) is always high-rank. Fortunately, the rank reduction procedure proposed in \cite{HP10} can be applied to find the optimal rank-1 solution. Let {\small$R_x = \text{rank}(\mathbf{X}_b)$} and {\small$\mathbf{X}_b = \mathbf{V}_x \mathbf{V}_x^H$} (for {\small$\mathbf{V}_x \in \mathbb{C}^{(r^2 + 1)\times R_x}$}). The optimal solution $\mathbf{X}_b$ is updated by
{\small
\begin{equation}
\label{EqRankReductionUpdate}
\mathbf{X}_{b,0} = \mathbf{V}_x \left(\mathbf{I} - 1/\delta_0 \mathbf{\Delta}\right) \mathbf{V}_x^H\,,
\end{equation}
}where $\mathbf{\Delta}$ is a $R_x$-by-$R_x$ Hermitian matrix satisfying
{\small
\begin{equation}
\label{EqSystem4Delta}
\text{Tr}\left\{\mathbf{V}_x^H \mathbf{B}_m \mathbf{V}_x \mathbf{\Delta} \right\} = 0\,, m = 2,3,4\,.
\end{equation}
}The coefficient $\delta_0$ in (\ref{EqRankReductionUpdate}) is calculated by $\delta_0 = \arg \max_{\{\delta_k \}_{k = 1}^{R_x}}|\delta_k|$, where $\delta_k$ denote the eigenvalues of $\mathbf{\Delta}$. The updated solution $\mathbf{X}_{b,0}$ (whose rank is at least one less than $\text{rank}(\mathbf{X}_b)$) preserves the primal feasibility and the complementary slackness such that it is optimal for the original problem\cite{HP10}. The optimal rank-1 solution can be found by repeating (\ref{EqRankReductionUpdate}) and (\ref{EqSystem4Delta}). Then, the optimal $\mathbf{b}^\star$ can be extracted from the rank-1 $\mathbf{X}_{b,0}$.

\subsection{Convergence of the Iterative Algorithm}
\begin{algorithm}
\caption{{\small The proposed iterative algorithm}\label{AlgIterAlgo}}
\begin{algorithmic}[1]
{\small
\State \textbf{Initialize} $\mathbf{A}_0^{(0)}$, $\mathbf{W}^{(0)}$, $\mathbf{F}^{(0)}$, and $\mathbf{B}_S^{(0)}$; set $\kappa \gets 0$;
\Repeat
    \State Given $(\mathbf{W}^{(\kappa)}, \mathbf{F}^{(\kappa)}, \mathbf{B}_S^{(\kappa)})$, update $\mathbf{A}_0^{(\kappa + 1)}$ by calculating (\ref{EqOptimA_0});\label{AlgLine_SubproA_0}
    \State Given $(\mathbf{A}_0^{(\kappa + 1)}, \mathbf{F}^{(\kappa)}, \mathbf{B}_S^{(\kappa)})$, update $\mathbf{W}^{(\kappa + 1)}$ by calculating (\ref{EqOptimW});\label{AlgLine_SubproW}
    \State Given $(\mathbf{A}_0^{(\kappa + 1)}, \mathbf{W}^{(\kappa + 1)}, \mathbf{B}_S^{(\kappa)})$, obtain $\mathbf{f}^\star$ by solving problem (\ref{QCQPsubproF_WMSEProBsF}); $\mathbf{F}^{(\kappa + 1)} \gets \text{reshape}(\mathbf{f}^\star, r_R, r_R)$;\label{AlgLine_SubproF}
    \State Given $(\mathbf{A}_0^{(\kappa + 1)}, \mathbf{W}^{(\kappa + 1)}, \mathbf{F}^{(\kappa + 1)})$, obtain $\mathbf{X}_b^\star$ by solving problem (\ref{SDRsubproBs_WMSEProBsF}) with CVX;
    Perform the rank reduction procedure (i.e. Algorithm 1 in \cite{HP10}) for $\mathbf{X}_b^\star$ and obtain the optimal rank-1 $\mathbf{X}_b^\prime$; $\mathbf{b}^\star \gets [\mathbf{b}^\prime]^\star /t^\star$; $\mathbf{B}_S^{(\kappa + 1)} \gets \text{reshape}(\mathbf{b}^\star, r, r)$;
    $\kappa \gets \kappa + 1$;\label{AlgLine_SubproBs}
\Until{\left| C_{\text{iter}}(\mathbf{A}_0^{(\kappa)}, \mathbf{W}^{(\kappa)}, \mathbf{F}^{(\kappa)}, \mathbf{B}_S^{(\kappa)}) - C_{\text{iter}}(\mathbf{A}_0^{(\kappa)}, \mathbf{W}^{(\kappa)}, \mathbf{F}^{(\kappa)}, \mathbf{B}_S^{(\kappa - 1)}) \right|<\varepsilon}\label{EqIterStopCriteria}
}
\end{algorithmic}
\end{algorithm}

Algorithm \ref{AlgIterAlgo} shows the proposed iterative algorithm, where {\small$C_{\text{iter}}(\mathbf{A}_0, \mathbf{W}, \mathbf{F}, \mathbf{B}_S) = \text{Tr}\{ \mathbf{A}_0 \mathbf{E}(\mathbf{W},$ $\mathbf{F}, \mathbf{B}_S) \} - \log \det (\mathbf{A}_0)$}. In each iteration (from Lines 3 to 7), the above four subproblems are solved, rank reduction is performed and the stopping criterion is checked.
\begin{theorem}
\label{ConvergenceIterativeAlgorithm}
The iterative algorithm as shown in Algorithm \ref{AlgIterAlgo} converges, as $\kappa$ tends to infinity.
\end{theorem}
\begin{IEEEproof}
Let {\small$\mathbf{x}^{(\kappa)} \triangleq (\mathbf{x}_1^{(\kappa)}, \ldots, \mathbf{x}_4^{(\kappa)})$} denote the sequence of the minimizers at the $\kappa$\,th iteration, where {\small$\mathbf{x}_1 \triangleq \text{vec}(\mathbf{A}_0)^T$}, {\small$\mathbf{x}_2 \triangleq \text{vec}(\mathbf{W})^T$}, {\small$\mathbf{x}_3 \! \triangleq \! \text{vec}(\mathbf{F})^T$}, and {\small$\mathbf{x}_4 \! \triangleq \! \text{vec}(\mathbf{B}_S)^T$}. Let {\small$\mathbf{y}_i^{(\kappa + 1)} \triangleq (\mathbf{x}_1^{(\kappa + 1)}, \ldots, \mathbf{x}_i^{(\kappa + 1)}, \mathbf{x}_{i+1}^{(\kappa)},\ldots, \mathbf{x}_4^{(\kappa)})$}. Since the subproblems (\ref{EqOptimA_0}), (\ref{EqOptimW}), and (\ref{QCQPsubproF_WMSEProBsF}) are convex and {\small$\mathbf{B}_S^{(\kappa + 1)}$} is the global optimal solution of problem (\ref{QCQPsubproBs_WMSEProBsF}), it is shown that
{\small
\begin{equation}
\label{EqC_iterMonoDecrease}
C_{\text{iter}}(\mathbf{x}^{(\kappa)}) \geq C_{\text{iter}}(\mathbf{y}_1^{(\kappa + 1)}) \geq C_{\text{iter}}(\mathbf{y}_2^{(\kappa + 1)}) \geq C_{\text{iter}}(\mathbf{y}_3^{(\kappa + 1)}) \geq C_{\text{iter}}(\mathbf{x}^{(\kappa + 1)})\,.
\end{equation}
}Thus, $C_{\text{iter}}(\mathbf{x}^{(\kappa)})$ monotonically decreases as $\kappa$ increases. Additionally, $C_{\text{iter}}(\cdot)$ is lower-bounded. Hence, $C_{\text{iter}}(\mathbf{x}^{(\kappa)})$ converges. Note that the stopping criterion of Algorithm \ref{AlgIterAlgo} is related to the convergence of $C_{\text{iter}}(\cdot)$ but not the convergence of minimizers as in \cite{BH03, Bertsekas99}. Therefore, we conclude that Algorithm \ref{AlgIterAlgo} converges.
\end{IEEEproof}

\begin{theorem}
\label{ConvergenceLimPointIterativeAlgorithm}
Suppose that tie-breaking strategies\cite{BH03} are included in solving problems (\ref{QCQPsubproF_WMSEProBsF}) and (\ref{SDRsubproBs_WMSEProBsF}), as well as the rank reduction procedure, such that $\mathbf{f}^\star$, $\mathbf{X}_b^\star$ and the rank-1 $\mathbf{X}_{b,0}$ are uniquely obtained. Then, the sequences {\small$\{ (\text{vec}(\mathbf{A}_0^{(\kappa)})^T, \text{vec}(\mathbf{W}^{(\kappa)})^T, \text{vec}(\mathbf{F}^{(\kappa)})^T, \text{vec}(\mathbf{B}_S^{(\kappa)})^T) \}_{\kappa = 0}^{\infty}$} converge to a unique limit point.
\end{theorem}
\begin{IEEEproof}
A tie-breaking strategy is a rule to select a solution from multiple solutions, e.g. to achieve the unique solution to problem (\ref{QCQPsubproF_WMSEProBsF}), the term $\mathcal{N}(\mathbf{A}_1)$ in (\ref{Eq_CloFormSolu_QCQPsubproF_constInactive}) can be omitted to yield the unique closed-form solution. To prove the theorem, we show that $\mathbf{x}^{(\kappa)}$ and $\mathbf{x}^{(\kappa + 1)}$ converge to the same limit point by contradiction \cite{Bertsekas99, Tseng01}. For details, please see Appendix \ref{Appendix_ConvergenceLimPointIterativeAlgorithm}.
\end{IEEEproof}

\begin{theorem}
\label{LimUnNCEstationary}
The limit point in Theorem \ref{ConvergenceLimPointIterativeAlgorithm} is not necessarily a stationary point of problem (\ref{WMSEProBsF}).
\end{theorem}
\begin{IEEEproof}
See Appendix \ref{Appendix_LimUnNCEstationary} for details.
\end{IEEEproof}

As a summary, Theorem \ref{ConvergenceIterativeAlgorithm} indicates that Algorithm \ref{AlgIterAlgo} can always converge, when the stopping criterion is designed as the difference of the objective functions (as shown in Line \ref{EqIterStopCriteria}), although the solution to each subproblem may not be unique. Alternatively, the criterion can also be related to the convergence of the minimizers \cite{BH03, Bertsekas99} such as
{\small
\begin{equation}
\label{EqConvLimCriterion}
\|\mathbf{x}^{(\kappa + 1)} - \mathbf{x}^{(\kappa)}\|/\|\mathbf{x}^{(\kappa + 1)}\| < \varepsilon^\prime\,.
\end{equation}
}Intuitively, if a subproblem has multiple global solutions, the minimizer may not converge. Theorem \ref{ConvergenceLimPointIterativeAlgorithm} illustrates that with such a criterion, the algorithm still converges (i.e. the minimizer converges) provided tie-breaking strategies are applied. Replacing Line \ref{EqIterStopCriteria} in Algorithm \ref{AlgIterAlgo} with (\ref{EqConvLimCriterion}), to make Algorithm \ref{AlgIterAlgo} converge, in the $\kappa$\,th iteration, (\ref{QCQPsubproF_WMSEProBsF}) can be solved by a numerical algorithm (e.g. an optimization solver) with a uniquely specified initial point (of $\mathbf{f}$) given {\small $(\mathbf{A}_1^{(\kappa)}, \mathbf{a}_1^{(\kappa)}, \mathbf{A}_2^{(\kappa)}, C_f^{(\kappa)})$}. The uniqueness means that if {\small $(\mathbf{A}_1^{(\kappa)}, \mathbf{a}_1^{(\kappa)}, \mathbf{A}_2^{(\kappa)}, C_f^{(\kappa)}) = (\mathbf{A}_1^{(\kappa^\prime)}, \mathbf{a}_1^{(\kappa^\prime)}, \mathbf{A}_2^{(\kappa^\prime)}, C_f^{(\kappa^\prime)})$} at two iterations $\kappa$ and $\kappa^\prime$, the initial points must be identical in those two iterations. Therefore, $\mathbf{f}^\star$ can be uniquely attained for a specific (\ref{QCQPsubproF_WMSEProBsF}). Similarly, to uniquely achieve $\mathbf{X}_b^\star$ for a specific (\ref{SDRsubproBs_WMSEProBsF}), a unique initial point of $\mathbf{X}_b$ should be specified for the specific $(\mathbf{B}_m, C_b, P_S)$. Then, in the rank reduction procedure, solving the system of (\ref{EqSystem4Delta}) numerically with a uniquely specified initial point of $\Delta$ for the specific $(\mathbf{V}_x, \mathbf{B}_m)$, a unique rank-1 $\mathbf{X}_{b,0}^\star$ can be finally obtained. Thus, $\mathbf{X}_{b,0}^\star$ is uniquely attained for a specific problem of (\ref{SDRsubproBs_WMSEProBsF}). In our implementation, CVX is exploited for (\ref{QCQPsubproF_WMSEProBsF}) and (\ref{SDRsubproBs_WMSEProBsF}), where the default solver SDPT3 solves a specific problem with a uniquely specified initial point\cite{TTT03}. The system of (\ref{EqSystem4Delta}) is solved by the fsolve function in MATLAB. Simulation results confirm the convergence.

\section{Simplified EFA Relaying Algorithms}
\label{SecHPM-PLM_Relaying}
To avoid the relatively high complexity caused by the matrices optimization\footnote{The proposed Algorithm \ref{AlgIterAlgo} solves four subproblems, among which three subproblems can yield closed-form solutions. The complexity of the IPA for solving (\ref{SDRsubproBs_WMSEProBsF}) is upper bounded by $\mathcal{O}(1)\left(7 + 2(r^2+1)\right)^{1/2} (r^2+1)^2 \left(5(r^2+1)^4 + 8(r^2+1)^3 +12(r^2+1)^2 +24\right)$ \cite{VBS09, BN01}. The number of iterations consumed by the following rank reduction procedure is upper bounded by $r^2 + 1$.}, considering the scenario\footnote{In the scenario $r < r_R$, precoder at $S$ can be the RSV (corresponding to non-zero singular values) of $\mathbf{H}_{R,S}$. The receiver at $R$ can be the conjugate transpose of the corresponding left singular vectors. The precoder at $D$ can be the linear combination of the RSV of $\mathbf{H}_{R, D}$ such that the EF leakage in ID can be close to a certain vector lying in the null space of $\mathbf{H}_{R,S}$. By this means, the power consumption of the retransmitted EF leakage can be well controlled. For the space constraint, here we do not discuss this scenario, while the case of $r = r_R$ is more non-trivial.} $r = r_R$, this section proposes the simplified EFA schemes. Specifically, by taking the singular value decomposition (SVD) of $\mathbf{H}_{D, R}$ and the joint decomposition of {\small$\mathbf{\tilde{H}}_{R,S} = \mathbf{H}_{R,S} \mathbf{Q}_S^{1/2}$} (i.e. the $S$-to-$R$ effective channel) and $\mathbf{H}_{R, D}$ based on the HPM-PLM strategy (which is discussed in Section \ref{SecHarvestedPowerMaximizationPowerLeakageMinimization}), the arguments (matrices) in $\det(\cdot)$ and $\text{Tr}(\cdot)$ in (\ref{ProblemP1}) can be diagonalized, such that the matrices optimization can be simplified to the power optimization.

\subsection{Channel Diagonalization}
\label{SecChannelDiagnalization}
\subsubsection{Structure of Relay Matrix}
Take the SVD of {\small$\mathbf{H}_{D, R} = \mathbf{U}_{D\!,R}\mathbf{\Sigma}_{D\!,R}\mathbf{V}_{D\!,R}^H$} and the SVD of {\small$\mathbf{\tilde{H}}_{R,S}\! =\! \mathbf{\tilde{U}}_{R,S} \mathbf{\tilde{\Sigma}}_{R,S}\mathbf{\tilde{V}}_{R,S}^H$}. Applying the matrix inversion lemma to (\ref{C_NonIF_P1}) yields {\small$1/2\cdot\log\det (\mathbf{I} + \frac{(1 - \rho)}{\sigma_n^2} \mathbf{\tilde{\Sigma}}_{R,S} (\mathbf{I} \! - \!(\mathbf{I}\! +\! \mathbf{\tilde{U}}_{R,S}^H \mathbf{F}^H \mathbf{H}_{D,R}^H \mathbf{H}_{D,R} \mathbf{F} \mathbf{\tilde{U}}_{R,S} )^{-1}) \mathbf{\tilde{\Sigma}}_{R,S} )$},
where the matrix between the two {\small$\mathbf{\tilde{\Sigma}}_{R,S}$} equals a positive semidefinite matrix {\small$\mathbf{\tilde{U}}_{R,S}^H \mathbf{F}^H\mathbf{H}_{D,R}^H(\mathbf{I}\! +\! \mathbf{H}_{D,R} \mathbf{F} \mathbf{F}^H \mathbf{H}_{D,R}^H)^{-1} \mathbf{H}_{D,R} \mathbf{F} \mathbf{\tilde{U}}_{R,S}$}. Hence, the matrix in $\det(\cdot)$ is positive-definite. According to Hadamard's inequality \cite{CT91}, the above $\log\det(\cdot)$ is maximized provided {\small$\mathbf{\tilde{U}}_{R,S}^H \mathbf{F}^H \mathbf{V}_{D,R} \mathbf{\Sigma}_{D,R}^2 \mathbf{V}_{D,R}^H \mathbf{F} \mathbf{\tilde{U}}_{R,S}$} is diagonal. Hence, {\small$\mathbf{F}\!=\!\mathbf{V}_{D,R} \mathbf{\Sigma}_F \mathbf{\tilde{U}}_{R,S}^H$} for {\small$\mathbf{\Sigma}_F\!\in\!\mathbb{C}^{r \times r}$} and the argument of the $\det(\cdot)$ in (\ref{C_NonIF_P1}) is diagonalized. Note that in the simplified EFA relaying, to maximize (\ref{C_NonIF_P1}) with diagonalized argument, all the power harvested at $R$ is used for forwarding (i.e. the inequality in (\ref{FwdPowerConst_1P1}) is converted to an equality). The structure of {\small$\mathbf{F}$} indicates that $R$ couples a given receive eigenmode of {\small$\mathbf{\tilde{U}}_{R,S}$} with a given transmit eigenmode of {\small$\mathbf{V}_{D,R}$} with an amplification factor given by the corresponding diagonal entry of {\small$\mathbf{\Sigma}_F$}.

\subsubsection{Maximize Harvested Power and Minimize Power Leakage}
\label{SecHarvestedPowerMaximizationPowerLeakageMinimization}
Take the eigenvalue decompositions (EVD) of {\small$\mathbf{Q}_D \!=\!\mathbf{V}_D \!\mathbf{\Sigma}_D^2 \!\mathbf{V}_D^H$} and {\small$\mathbf{H}_{R,D}\! \mathbf{Q}_D  \mathbf{H}_{R,D}^H\! = \!\mathbf{\tilde{U}}_{R,D}\! \mathbf{\tilde{\Sigma}}_{R,D}^2 \! \mathbf{\tilde{U}}_{R,D}^H$}. Recall that the inequality in (\ref{FwdPowerConst_1P1}) has been converted to an equality. With the SVD of {\small$\mathbf{H}_{R,D}$}, rearranging (\ref{FwdPowerConst_1P1}) yields
{\small
\begin{IEEEeqnarray}{cl}
\label{FwdPowerConst_2}
&\text{Tr}\left\{(1 - \rho) \mathbf{\Sigma}_F^H \mathbf{\Sigma}_F \mathbf{\tilde{\Sigma}}_{R,S}^2 + \sigma_n^2\mathbf{\Sigma}_F^H \mathbf{\Sigma}_F - \rho \mathbf{\tilde{\Sigma}}_{R,S}^2\right\} \IEEEyessubnumber \label{FwdPowerConst_2_1} \\
{}={}&\text{Tr}\left\{ \! \left(\rho \mathbf{I}\! - \! (1 \! - \! \rho)\mathbf{\Sigma}_F^H \mathbf{\Sigma}_F\right)\! \left(\mathbf{\tilde{U}}_{R,S}^H \mathbf{V}_{D,R}^{\ast} \mathbf{\Sigma}_{D,R} \mathbf{U}_{D,R}^T \mathbf{V}_D \mathbf{\Sigma}_D^2 \mathbf{V}_D^H \mathbf{U}_{D,R}^{\ast} \mathbf{\Sigma}_{D,R} \mathbf{V}_{D,R}^T \mathbf{\tilde{U}}_{R,S}\right) \! \right\} \IEEEyessubnumber \label{FwdPowerConst_2_2}\\
{}={}& \! \text{Tr} \! \left\{ \! \left(\rho \mathbf{I} \! - \! (1 \! - \! \rho)\mathbf{\Sigma}_F^H \mathbf{\Sigma}_F \! \right) \! \left( \! \mathbf{\tilde{U}}_{R,S}^H \! \mathbf{\tilde{U}}_{R,D} \! \mathbf{\tilde{\Sigma}}_{R,D}^2 \! \mathbf{\tilde{U}}_{R,D}^H \! \mathbf{\tilde{U}}_{R,S} \! \right) \! \right\}\!. \IEEEyessubnumber \label{FwdPowerConst_2_3}
\end{IEEEeqnarray}
}Eq. (\ref{FwdPowerConst_2_3}) describes the difference between the power of the EF harvested at the EH receiver and the EF leaking into the ID receiver. Thus, a strategy can be proposed to maximize the harvested power at $R$ and minimize the power leakage. Recall that the harvested power has been maximized by letting {\small$\mathbf{Q}_D \! =\! P_D [\mathbf{U}_{D,R}^{\ast}]_{\text{max}} [\mathbf{U}_{D,R}^{\ast}]_{\text{max}}^H$}, such that {\small$\rho \text{Tr}\{ \! \mathbf{H}_{R, D} \mathbf{Q}_D \mathbf{H}_{R, D}^H \!\} \! = \!\rho P_D\lambda_{D,R,\text{max}}$}. To minimize the power leakage, the EF leaking into the ID receiver should be paired with the minimum amplification coefficient, such that the power of the retransmitted leakage equals {\small$(1 - \rho)\lambda_{f,\text{min}} P_D \lambda_{D,R,\text{max}}$}, where $\lambda_{f,\text{min}}$ denotes the minimum diagonal entry of {\small$\mathbf{\Sigma}_F^H \mathbf{\Sigma}_F$}. With the HPM-PLM strategy, (\ref{FwdPowerConst_2_3}) should equal {\small$(\rho - (1 - \rho)\lambda_{f,\text{min}}) P_D \lambda_{D,R,\text{max}}$}, which is shown to be an upper bound of (\ref{FwdPowerConst_2_3}) by applying Lemma II.1 in \cite{Lasserre95}. To make (\ref{FwdPowerConst_2_3}) equal to the upper bound, {\small$\mathbf{\tilde{U}}_{R,S}^H \mathbf{V}_{D,R}^{\ast} = \mathbf{P_\pi}$} in (\ref{FwdPowerConst_2_2}), where $\mathbf{P_\pi}$ permutates the unique non-zero diagonal entry {\small$P_D \lambda_{D,R,\text{max}}$} in the diagonal matrix {\small$\mathbf{\Sigma}_{D,R} \mathbf{U}_{D,R}^T \mathbf{V}_D \mathbf{\Sigma}_D^2 \mathbf{V}_D^H \mathbf{U}_{D,R}^{\ast} \mathbf{\Sigma}_{D,R}$} to the same position as {\small$\rho - (1 - \rho)\lambda_{f,\text{min}}$} in {\small$\rho \mathbf{I} \! - \! (1 \! - \! \rho)\mathbf{\Sigma}_F^H \mathbf{\Sigma}_F$}. By this means, (\ref{FwdPowerConst_2_3}) achieves the upper bound, and the matrices in the traces of (\ref{FwdPowerConst_1P1}) is diagonalized. In summary, the argument of the $\det(\cdot)$ in (\ref{C_NonIF_P1}) is diagonalized with the decomposed {\small$\mathbf{H}_{D,R}$} and the structure of {\small$\mathbf{F}$}. The argument of the traces in (\ref{FwdPowerConst_1P1}) is diagonalized with the HPM-PLM strategy, i.e. the rank-1 {\small$\mathbf{Q}_D$} and {\small$\mathbf{\tilde{U}}_{R,S}\!=\!\mathbf{V}_{D,R}^{\ast}\mathbf{P}_{\mathbf{\pi}}^T$}. Since {\small$\mathbf{Q}_S \! = \! \mathbf{H}_e \mathbf{\tilde{\Sigma}}_{R,S}^2 \mathbf{H}_e ^{H}$} where {\small$\mathbf{H}_e \! = \! (\mathbf{\tilde{U}}_{R,S}^H \mathbf{H}_{R, S})^{-1}$}, (\ref{SrcPowerConst_1P1}) becomes {\small$\text{Tr}\{\mathbf{Q}_S\} \! = \! \sum_{m = 1}^r \|\mathbf{h}_{e,m}\|^2 \tilde{\lambda}_{R,S,m} \leq P_S$}, where {\small$\mathbf{h}_{e,m} = [\mathbf{H}_e]_m$} and {\small$\tilde{\lambda}_{R,S,m}$} denotes the $m$\,th diagonal entry of {\small$\mathbf{\tilde{\Sigma}}_{R,S}^2$}. Hence, problem (\ref{ProblemP1}) reduces to the power optimization.

\subsubsection{Discussion}
Substituting {\small$\mathbf{\tilde{U}}_{R,S}^H  = \mathbf{P_\pi}\mathbf{V}_{D,R}^T$} into the structure of {\small$\mathbf{F}$}, we have {\small$\mathbf{F}  = \mathbf{V}_{D,R} \mathbf{\Sigma}_F \mathbf{P}_{\mathbf{\pi}} \cdot \mathbf{V}_{D,R}^T$}. Namely, the RSV of {\small$\mathbf{F}$} matches the left singular vectors (LSV) of {\small$\mathbf{\tilde{H}}_{R,S}$} and a permutation of the LSV of {\small$\mathbf{H}_{D, R}$}. Thus, the power of $\mathbf{y}_D$ is given by
{\small
\begin{equation}
\mathcal{E}\{\|\mathbf{y}_D\|^2\} = \text{Tr} \{(1\! -\! \rho) \mathbf{\Sigma}_{D,R}^2 \mathbf{\Sigma}_F\mathbf{\Sigma}_F^H (\mathbf{\tilde{\Sigma}}_{R,S}^2\! +\! \mathbf{P}_{\mathbf{\pi}} \mathbf{\Sigma}_{D,R}^2 \mathbf{U}_{D,R}^T \mathbf{Q}_D \mathbf{U}_{D,R}^\ast \mathbf{P}_{\mathbf{\pi}}^T)\! + \! (\mathbf{\Sigma}_{D,R}^2 \mathbf{\Sigma}_F\mathbf{\Sigma}_F^H \! + \! \mathbf{I})\sigma_n^2\}\,,\label{EqPwr_yD}
\end{equation}
}where {\small$\mathbf{U}_{D,R}^T\mathbf{Q}_D\mathbf{U}_{D,R}^\ast$} is diagonal. In (\ref{EqPwr_yD}), because {\small$\mathbf{\tilde{H}}_{R,S}$} and {\small$\mathbf{H}_{R,D}$} share the same (but permutated) LSV, the channel power gains of the effective $S$-to-$R$ and $D$-to-$R$ channels in phase 1 are overlapped at the ID receiver, i.e. {\small$(1\!-\!\rho) (\mathbf{\tilde{\Sigma}}_{R,S}^2 \! + \! \mathbf{P}_{\mathbf{\pi}} \mathbf{\Sigma}_{D,R}^2  \mathbf{U}_{D,R}^T \mathbf{Q}_D \mathbf{U}_{D,R}^\ast  \mathbf{P}_{\mathbf{\pi}}^T)$}. Although the retransmitted energy flow can be canceled at $D$ (i.e. {\small$\mathbf{P}_{\mathbf{\pi}} \mathbf{\Sigma}_{D,R}^2 \mathbf{U}_{D,R}^T \mathbf{Q}_D\mathbf{U}_{D,R}^\ast \mathbf{P}_{\mathbf{\pi}}^T$} in (\ref{EqPwr_yD}) is eliminated), the overlapped channel power gains in phase 1 still impact the rate, because the EF leakage is retransmitted and consume power. Denote the diagonal entries of {\small$\mathbf{\tilde{\Sigma}}_{R,S}^2$} and {\small$(1 - \rho)\mathbf{P}_{\mathbf{\pi}} \mathbf{\Sigma}_{D,R}^2 \mathbf{U}_{D,R}^T \mathbf{Q}_D \mathbf{U}_{D,R}^\ast \mathbf{P}_{\mathbf{\pi}}^T$} as {\small$\mathbf{\tilde{\lambda}}_{R,S} \triangleq [\tilde{\lambda}_{R,S,1}, \ldots, \tilde{\lambda}_{R,S,r}]^T$} and {\small$\mathbf{\beta} \triangleq [\beta_1, \ldots, \beta_r]^T$} (where the unique non-zero {\small$\beta_m\!=\!(1\!-\!\rho)P_D \lambda_{D,R,\text{max}} \triangleq c$}), respectively. In the diagonalized relay power constraint (\ref{FwdPowerConst_1P1}) (i.e. the following (\ref{EqEqConstP3a})), the diagonal entries of {\small$\mathbf{\Sigma}_F\mathbf{\Sigma}_F^H$}, denoted as {\small$\mathbf{\lambda}_f \triangleq [\lambda_{f,1}, \ldots, \lambda_{f,r}]^T$}, are multiplied by the entries of the overlapped channel power gains, i.e. {\small$(1 - \rho)\tilde{\lambda}_{R,S,m} + \beta_m$} for $m = 1,\ldots, r$. Thus, the pairings of the diagonal entries of {\small$\mathbf{\Sigma}_{D,R}^2$} (which is also coupled with {\small$\mathbf{\Sigma}_F\mathbf{\Sigma}_F^H$}) and the overlapped channel power gains in phase 1 affect the optimization of {\small$\mathbf{\lambda}_f$} and thereby the rate. Additionally, in phase 1, the value of each {\small$(1 - \rho)\tilde{\lambda}_{R,S,m} + \beta_m$} is affected by {\small$\mathbf{P}_{\mathbf{\pi}}$} (which determines the pairing of each {\small$\tilde{\lambda}_{R,S,m}$} and {\small$\beta_m$}).

\subsection{Joint Power Allocation Optimization}
\label{SecJointOptimization}
To make further calculation and analysis tractable, the power optimization problem only maximizes the achievable rate at high receive SNR. Substituting the previous channel decompositions into problem (\ref{ProblemP1}), the problem can be reformulated as
{\small
\begin{IEEEeqnarray}{cl}
\label{ProblemP2}
\min_{\mathbf{\lambda}_f,\mathbf{\tilde{\lambda}}_{R,S}}\, & -\sum_{m = 1}^{r} \log \left(\frac{(1 - \rho) \tilde{\lambda}_{R,S,m} \lambda_{f,m} \lambda_{D,R,m}}{\sigma_n^2\left(1 + \lambda_{f,m} \lambda_{D,R,m}\right)}\right) \IEEEyessubnumber \label{C_NonIF_Scalar_P2} \\
\text{s.t.} & \lambda_{f,1}, \lambda_{f,2}, \ldots, \lambda_{f,r} > 0\,, \IEEEyessubnumber \label{EqldaFineqConst}\\
& \tilde{\lambda}_{R,S,1}, \tilde{\lambda}_{R,S,2}, \ldots, \tilde{\lambda}_{R,S,r} > 0\,, \IEEEyessubnumber \label{EqldaRSineqConst} \\
& \text{Tr}\{\mathbf{Q}_S\} = \sum_{m = 1}^r \|\mathbf{h}_{e,m}\|^2 \tilde{\lambda}_{R,S,m} \leq P_S,\IEEEyessubnumber \label{EqSrcPwrConst} \\
& \sum_{m = 1}^r \lambda_{f,m} \left( (1 - \rho)\tilde{\lambda}_{R,S,m} + \sigma_n^2 + \beta_m\right)\!=\!\sum_{m = 1}^r \rho \tilde{\lambda}_{R,S,m} + \rho P_D \lambda_{D,R,\text{max}} \,, \IEEEyessubnumber \label{EqEqConstP3a}
\end{IEEEeqnarray}
}where $\beta_m$ is constrained by $\beta_m\!=\!c$ for $m\!=\!\text{index}(\lambda_{f,\text{min}})$ (where $\text{index}(\!\lambda_{f,\text{min}}\!)$ returns the index of $\lambda_{f,\text{min}}$); otherwise, $\beta_m =0$. Problem (\ref{ProblemP2}) is not convex due to the non-affine (\ref{EqEqConstP3a}). Then, problem (\ref{ProblemP2}) is solved by performing power optimizations for $R$ and $D$ alternatively.

\subsubsection{Relay Optimization with Fixed Source Power Allocation}
\label{SecOptimRelayPwrAllocWithFixedSrcPwrAlloc}
With given $\tilde{\lambda}_{R,S,m}$, the power optimization problem at $R$ is formulated as
{\small
\begin{IEEEeqnarray}{cl}
\label{ProblemP3a}
\max_{\mathbf{\lambda}_f}\,&  \sum_{m = 1}^{r} \log \left( \frac{(1 - \rho) \tilde{\lambda}_{R,S,m} \lambda_{f,m} \lambda_{D,R,m}}{\sigma_n^2\left(1 + \lambda_{f,m} \lambda_{D,R,m}\right)}\right) \\
\text{s.t.} & \text{(\ref{EqldaFineqConst}) and (\ref{EqEqConstP3a})}\,.\nonumber
\end{IEEEeqnarray}
}The challenge in solving problem (\ref{ProblemP3a}) is that $c$ of $\beta_m$ is required to be paired with $\lambda_{f,\text{min}}$, but the position of $\lambda_{f,\text{min}}$ in $\mathbf{\lambda}_f$ is unknown before solving the problem; the rate is affected by the pairings of the elements of {\small$\mathbf{\lambda}_{D,R}\triangleq [\lambda_{D,R,1}, \ldots, \lambda_{D,R,r}]^T$} and $\mathbf{\tilde{\lambda}}_{R,S}$ (i.e. the pairings of the eigenmodes of the forwarding channel {\small$\mathbf{H}_{D,R}$} and the eigenmodes of the effective $S$-to-$R$ channel {\small$\mathbf{\tilde{H}}_{R,S}$}), even if the constraint on $c$ is relaxed and $\beta_m$ is fixed. To avoid the high complexity of searching the best pairings, we then reveal that the pairing issues can be solved by ordering operations. Firstly, we relax the constraint on $c$, i.e. $\mathbf{P}_{\mathbf{\pi}}$ becomes an arbitrary permutation matrix {\small$\mathbf{\tilde{P}}_{\mathbf{\pi}}$}, and assume columns of {\small$\mathbf{V}_{D,R}$} are arranged in certain orders, such that the elements of {\small$\mathbf{\lambda}_{D,R}$} and {\small$\mathbf{\tilde{\lambda}}_{R,S}$} are paired in certain ways and $c$ is paired with a certain {\small$\tilde{\lambda}_{R,S,m}$}. Problem (\ref{ProblemP3a}) then becomes a convex problem regardless of the pairing issues. By analyzing KKT conditions, a closed-form solution can be obtained by
{\small
\begin{IEEEeqnarray}{l}
\label{EqClosedSoluP3a}
\lambda^{\star}_{f,m} = - \frac{1}{2\lambda_{D,R,m}} + \frac{1}{2} \! \sqrt{\frac{1}{\lambda^2_{D,R,m}} \! + \! \frac{4}{\nu^{\star}\lambda_{D,R,m}\left(\!(1 \! - \! \rho)\tilde{\lambda}_{R,S,m} \! + \! \sigma_n^2 \! + \! \beta_m \right)}},
\end{IEEEeqnarray}
}where $\nu^{\star}$ denotes the Lagrange multiplier for (\ref{EqEqConstP3a}) and is greater than 0 (because {\small$\lambda^{\star}_{f,m} > 0$}). It can be calculated by solving {\small$\sum^r_{m = 1} \lambda^{\star}_{f,m}\left((1\!-\!\rho)\tilde{\lambda}_{R,B,m}\!+\!\sigma_n^2 + \beta_m \right)\!=\!\rho P_D \lambda_{D,R,\text{max}} + \sum^r_{m = 1}\rho\tilde{\lambda}_{R,S,m}$} with bisection. Based on (\ref{EqClosedSoluP3a}), two lemmas are revealed. For notational simplicity, let {\small$z_m \triangleq (1 - \rho)\tilde{\lambda}_{R,S,m} + \sigma_n^2 + \beta_m$}, {\small$\mathbf{z} \triangleq [z_1, \ldots, z_r]^T$}; {\small$l_m \triangleq (1 - \rho)\tilde{\lambda}_{R,S,m} + \sigma_n^2$}, {\small$\mathbf{l} \triangleq [l_1, \ldots, l_r]^T$}.
\begin{lemma}
\label{Lemma1}
Suppose that the elements in $\pi_1(\mathbf{z})$ are arranged in the same order as another permutation $\pi_2(\mathbf{z})$ except that $z_i$ and $z_j$ (where $z_i \leq z_j$ for $i < j$) in $\pi_1(\mathbf{z})$ are swapped in $\pi_2(\mathbf{z})$. Namely, $z_i$ and $z_j$ in $\pi_1(\mathbf{z})$ are respectively paired with $\lambda_{D,R,p}$ and $\lambda_{D,R,q}$ (where $\lambda_{D,R,p} \leq \lambda_{D,R,q}$ for $p < q$), while $z_j$ and $z_i$ in $\pi_2(\mathbf{z})$ are respectively paired with $\lambda_{D,R,p}$ and $\lambda_{D,R,q}$. Then, the value of the objective function of (\ref{ProblemP3a}) with $\pi_1(\mathbf{z})$ is no less than that with $\pi_2(\mathbf{z})$.
\end{lemma}
\begin{IEEEproof}
This lemma is proved by respectively substituting $\pi_1(\mathbf{z})$ and $\pi_2(\mathbf{z})$ into (\ref{EqClosedSoluP3a}) with $\mathbf{\lambda}_{D,R}$ and comparing the values of the objective function of (\ref{ProblemP3a}). See Appendix \ref{Appendix_ProofLemma1} for details.
\end{IEEEproof}
Lemma \ref{Lemma1} addresses the pairings of the transmit eigenmodes of $\mathbf{F}$ (i.e. $\mathbf{V}_{D,R}$) and the overlapped channel power gains. With fixed pairings of $\tilde{\lambda}_{R,S,m}$ and $\beta_m$ for $m = 1\ldots r$, the values of the entries of the overlapped channel power gains, i.e. $(1-\rho)\tilde{\lambda}_{R,S,m} + \beta_m$ for $m = 1\ldots r$, are fixed. Lemma \ref{Lemma1} reveals that, for two pairs of the transmit eigenmodes of $\mathbf{V}_{D,R}$ and the overlapped channel power gains (while other pairings are fixed), the strongest eigenmode of $\mathbf{V}_{D,R}$ and the strongest overlapped channel power gain should be paired together. The following Lemma \ref{Lemma2} addresses the pairings of the channel power gains of the effective $S$-to-$R$ channel and the non-zero channel power gain of the effective $D$-to-$R$ channels, i.e. the pairings of $\tilde{\lambda}_{R,S,m}$ and $c$ (i.e. the unique non-zero $\beta_m$) for $m = 1 \ldots r$. It is shown that, for two channel power gains in $\tilde{\lambda}_{R,S,m}$, the strongest $\tilde{\lambda}_{R,S,m}$ should be paired with $c$.
\begin{lemma}
\label{Lemma2}
Assume two permutations $\pi_1(\mathbf{l})$ and $\pi_2(\mathbf{l})$. In $\pi_1(\mathbf{l})$, positions of $l_i$ and $l_j$ follows that $\min\{l_i + c, l_j\}$ and $\max\{l_i + c, l_j\}$ are respectively paired with $\lambda_{D,R,i}$ and $\lambda_{D,R,j}$ (where $i < j$, $l_i \leq l_j$ and $\lambda_{D,R,i} \leq \lambda_{D,R,j}$). In $\pi_2(\mathbf{l})$, positions of $l_i$ and $l_j$ follows that $\min\{l_i, l_j + c\}$ and $\max\{l_i, l_j + c\}$ are respectively paired with $\lambda_{D,R,i}$ and $\lambda_{D,R,j}$. Other pairings between $\lambda_{D,R,m}$ and $l_m$ (for $m \neq i,j$) in $\pi_1(\mathbf{l})$ are the same as $\pi_2(\mathbf{l})$. Then, the objective function of (\ref{ProblemP3a}) with $c$ paired with $l_j$ yields a higher value than that with $c$ paired with $l_i$.
\end{lemma}
\begin{IEEEproof}
Lemma \ref{Lemma2} is proved based on Lemma \ref{Lemma1}. The proof strategy is similar to Lemma \ref{Lemma1}. See Appendix \ref{Appendix_ProofLemma2} for details.
\end{IEEEproof}
\begin{proposition}
\label{Proposition1}
When the elements in $\mathbf{\tilde{\lambda}}_{R,S}$ and $\mathbf{\lambda}_{D,R}$ are arranged in increasing orders and the non-zero $\beta_m$ is paired with the maximum $\tilde{\lambda}_{R,S,m}$, the value of the objective function of (\ref{ProblemP3a}) is maximized and the optimal $\lambda^{\star}_{f,m}$ are arranged in a decreasing order.
\end{proposition}
\begin{IEEEproof}
Applying Lemma \ref{Lemma1} and Lemma \ref{Lemma2}, the orderings of $\mathbf{\tilde{\lambda}}_{R,S}$ and $\mathbf{\lambda}_{D,R}$ and the pairing of $\beta_m$ and $\tilde{\lambda}_{R,S,m}$ are proved by induction. The decreasing order of $\lambda^{\star}_{f,m}$ requires that {\small$\lambda^{\star}_{f,m} - \lambda^{\star}_{f,m+1} = - \frac{1}{2a_m} + \frac{1}{2}\sqrt{\frac{1}{a^2_m} + \frac{4}{\nu^{\star}a_m z_m}} - (- \frac{1}{2a_{m+1}} + \frac{1}{2}\sqrt{\frac{1}{a^2_{m+1}} + \frac{4}{\nu^{\star}a_{m+1} z_{m+1}}}) \geq 0$}, where $a_m = \lambda_{D,R,m}$. Rearranging the above inequality, the proof ends up showing $\nu^{\star} \geq - \frac{(a_m z_m - a_{m+1}z_{m+1})^2}{z_m(z_m - z_{m+1})z_{m+1}(a_m - a_{m+1})}$. The proved ordering and pairing show that $a_m \leq a_{m+1}$ and $z_m \leq z_{m+1}$. Since optimal solution (\ref{EqClosedSoluP3a}) is achieved only when $\nu^{\star} > 0$, there always exists $\lambda^{\star}_{f,m} \geq \lambda^{\star}_{f,m+1}$, i.e. $\lambda^{\star}_{f,m}$ are arranged in a decreasing order. So far, Proposition \ref{Proposition1} has been proved.
\end{IEEEproof}

Proposition \ref{Proposition1} illustrates that the constraint on $\beta_m$ (i.e. $c$ is paired with $\lambda_{f,\text{min}}$) can be safely relaxed. Following the ordering operation in Proposition \ref{Proposition1}, entries $(\rho - (1 - \rho)\lambda_{f,\text{min}})$ and $P_D \lambda_{D,R,\text{max}}$ are at lower-right corners of matrices {\small$\rho \mathbf{I} - (1 - \rho) \mathbf{\Sigma}_F^H \mathbf{\Sigma}_F$} and {\small$\mathbf{\Sigma}_{D,R} \mathbf{U}_{D,R}^T \mathbf{V}_D \mathbf{\Sigma}_D^2  \mathbf{V}_D^H  \mathbf{U}_{D,R}^{\ast} \mathbf{\Sigma}_{D,R}$} in (\ref{FwdPowerConst_2}), respectively. Hence, the permutation matrix {\small$\mathbf{\tilde{P}}_{\mathbf{\pi}} = \mathbf{I} = \mathbf{P}_{\mathbf{\pi}}$}.

\subsubsection{Source Optimization with Fixed Relay Power Allocation}
According to Proposition \ref{Proposition1}, $\lambda_{f,m}$ are arranged in a decreasing order, and $\text{index}(\lambda_{f,\text{min}}) =r$. Thus, the source power optimization problem is formulated as
{\small
\begin{IEEEeqnarray}{cl}
\label{ProblemP3b}
\min_{\mathbf{\tilde{\lambda}}_{R,S}}\,& - \! \sum_{m = 1}^{r} \log \left( \frac{(1 - \rho) \tilde{\lambda}_{R,S,m} \lambda_{f,m} \lambda_{D,R,m}}{\sigma_n^2\left(1 + \lambda_{f,m} \lambda_{D,R,m}\right)}\right) \IEEEyessubnumber \label{EqObjFuncP3b} \\
\text{s.t.} & 0 < \tilde{\lambda}_{R,S,1} \leq \tilde{\lambda}_{R,S,2} \leq \ldots \leq \tilde{\lambda}_{R,S,r}\,, \IEEEyessubnumber \label{EqIneqConst1P3b}\\
& \text{(\ref{EqSrcPwrConst}) and (\ref{EqEqConstP3a})}\,.\nonumber
\end{IEEEeqnarray}
}Problem (\ref{ProblemP3b}) is convex and can be solved by an optimization solver. Analytical solutions are still attractive due to its low complexity. The challenge in deriving a closed-form solution is the ordering constraint in (\ref{EqIneqConst1P3b}). However, when the ordering constraint in (\ref{EqIneqConst1P3b}) is relaxed, the output $\tilde{\lambda}_{R,S,m}^\star$ can still be in an increasing order if $\tilde{\lambda}_{R,S,m}$ in (\ref{EqSrcPwrConst}) are uniformly weighted (otherwise, the Lagrange multiplier for (\ref{EqSrcPwrConst}) would be non-uniformly weighted in the derived closed-form solution, which may violate the ordering of $\mathbf{\tilde{\lambda}}_{R,S}$). Therefore, problem (\ref{ProblemP3b}) can be simplified by respectively replacing constraints (\ref{EqIneqConst1P3b}) and (\ref{EqSrcPwrConst}) with $\tilde{\lambda}_{R,S,1}, \tilde{\lambda}_{R,S,2}, \ldots, \tilde{\lambda}_{R,S,r} > 0$ and $\sum_{m = 1}^r h_{e,\text{max}}^2 \tilde{\lambda}_{R,S,m} \leq P_S$ (where $h_{e,\text{max}}^2 = \max\{\|\mathbf{h}_{e,m}\|^2\}$). This simplified problem is referred to as the \emph{simplified source power optimization}. The simplified source power optimization is convex, and its KKT conditions are listed as follows.
{\small
\begin{IEEEeqnarray}{c}
\tilde{\lambda}^{\star}_{R,S,m} > 0\,,\quad m = 1,\ldots,r \IEEEyessubnumber \label{EqKKT1P3c}\\
\sum_{m = 1}^r \tilde{\lambda}^{\star}_{R,S,m} \leq P_S/h_{e,\text{max}}^2\,,\IEEEyessubnumber \label{EqKKT2P3c}\\
\sum_{m = 1}^r \left(\lambda_{f,m}(1 - \rho) - \rho\right)\tilde{\lambda}^{\star}_{R,S,m} = - \sum_{m = 1}^r (\sigma_n^2 + \beta_m)\lambda_{f,m} + \rho P_D \lambda_{D,R,\text{max}}\,,\IEEEyessubnumber \label{EqKKT3P3c}\\
\gamma^{\star}_{1,m} \geq 0\,,\, \gamma^{\star}_{1,m}\tilde{\lambda}^{\star}_{R,S,m} = 0\,,\IEEEyessubnumber \label{EqKKT5P3c}\\
\gamma^{\star}_2 \geq 0\,,\, \gamma^{\star}_2 \left( \sum_{m = 1}^r \tilde{\lambda}^{\star}_{R,S,m} - P_S/h_{e,\text{max}}^2 \right) = 0\,,\IEEEyessubnumber \label{EqKKT7P3c}\\
- 1/\tilde{\lambda}^{\star}_{R,S,m} - \gamma^{\star}_{1,m} + \gamma^{\star}_2 + \mu^{\star}\left(\lambda_{f,m}(1 - \rho) - \rho\right) = 0,\IEEEyessubnumber \label{EqKKT8P3c}
\end{IEEEeqnarray}
}where $\gamma^{\star}_{1,m}$, $\gamma^{\star}_2$, and $\mu^{\star}$ denote the optimal Lagrange multipliers. Eq. (\ref{EqKKT1P3c}) and (\ref{EqKKT5P3c}) reveal that $\gamma^{\star}_{1,m} = 0$. If $\gamma^{\star}_2 = 0$, according to (\ref{EqKKT8P3c}), it is obtained that
{\small
\begin{equation}
\label{EqClosedSolu1P3c}
\tilde{\lambda}^{\star}_{R,S,m} = 1/(\mu^{\star} \left(\lambda_{f,m}(1 - \rho) - \rho\right))\,,
\end{equation}
}where $\mu^{\star}$ is obtained by solving $r / \mu^{\star} = \rho P_D \lambda_{D,R,\text{max}} - \sum_{m = 1}^r (\sigma_n^2 + \beta_m)\lambda_{f,m}$. Since $\tilde{\lambda}^{\star}_{R,S,m} > 0$ $\forall m$ and $\mu^{\star}$ also conforms to (\ref{EqKKT2P3c}), (\ref{EqClosedSolu1P3c}) is obtained provided
{\small
\begin{equation}
\label{EqP3cSolutionCond1}
\begin{cases}
\lambda_{f,m}(1 - \rho) - \rho \gtrless 0\,,\quad \forall m\\
0 \lessgtr \frac{1}{\mu^{\star}} \lesseqgtr \frac{P_S/h_{e,\text{max}}^2}{\sum^r_{m = 1}\frac{1}{\lambda_{f,m}(1 - \rho) - \rho}}
\end{cases}
\end{equation}
}is satisfied. On the other hand, if $\lambda_3 > 0$, the optimal $\tilde{\lambda}^{\star}_{R,S,m}$ is achieved by
{\small
\begin{equation}
\label{EqClosedSolu2P3c}
\tilde{\lambda}^{\star}_{R,S,m} = 1/(\gamma^{\star}_2 + \mu^{\star} \left(\lambda_{f,m}(1 - \rho) - \rho\right))\,,
\end{equation}
}where $\gamma^{\star}_2$ and $\mu^{\star}$ can be obtained by solving the non-linear system composed of (\ref{EqKKT2P3c}) and (\ref{EqKKT3P3c}).

\begin{figure*}[t]
\begin{minipage}[t]{.48\textwidth}
\begin{algorithm}[H]
\caption{{\small EFA-S1}}\label{AlgTwoPhaseRelayingP3b}
\begin{algorithmic}[1]
{\small
\State \textbf{Initialize} $\mathbf{\lambda}_f^{(0)}$ and $\mathbf{\tilde{\lambda}}_{R,S}^{(0)}$
\Repeat
    \State Update $\mathbf{\lambda}_f^{(\kappa + 1)}$ by calculating (\ref{EqClosedSoluP3a});
    \State Update $\mathbf{\tilde{\lambda}}_{R,S}^{(\kappa + 1)}$ by solving (\ref{ProblemP3b}) with an optimization solver;
    \State $\kappa \gets \kappa + 1$;
\Until{\left|C(\mathbf{\lambda}_f^{(\kappa)}, \mathbf{\tilde{\lambda}}_{R,S}^{(\kappa)})\!-\!C(\mathbf{\lambda}_f^{(\kappa)}, \mathbf{\tilde{\lambda}}_{R,S}^{(\kappa - 1)})\right|\!<\!\epsilon}
}
\end{algorithmic}
\end{algorithm}
\end{minipage}
\hfill
\begin{minipage}[t]{.48\textwidth}
\begin{algorithm}[H]
\caption{{\small EFA-S2}}\label{AlgTwoPhaseRelayingP3c}
\begin{algorithmic}[1]
{\small
\State \textbf{Initialize} $\mathbf{\lambda}_f^{(0)}$ and $\mathbf{\tilde{\lambda}}_{R,S}^{(0)}$
\Repeat
    \State Update $\mathbf{\lambda}_f^{(\kappa + 1)}$ by calculating (\ref{EqClosedSoluP3a});
    \State \textbf{if} {\text{(\ref{EqP3cSolutionCond1}) is satisfied}} \textbf{then} update $\mathbf{\tilde{\lambda}}_{R,S}^{(\kappa + 1)}$ by calculating (\ref{EqClosedSolu1P3c});
    \State \textbf{else} update $\mathbf{\tilde{\lambda}}_{R,S}^{(\kappa + 1)}$ by calculating (\ref{EqClosedSolu2P3c});
    \EndIf
    \State $\kappa \gets \kappa + 1$;
\Until{\left|C(\mathbf{\lambda}_f^{(\kappa)}, \mathbf{\tilde{\lambda}}_{R,S}^{(\kappa)})\! -\! C(\mathbf{\lambda}_f^{(\kappa)}, \mathbf{\tilde{\lambda}}_{R,S}^{(\kappa - 1)})\right|\!<\!\epsilon}
}
\end{algorithmic}
\end{algorithm}
\end{minipage}
\end{figure*}
As a summary, the simplified EFA algorithms are outlined in Algorithms \ref{AlgTwoPhaseRelayingP3b} and \ref{AlgTwoPhaseRelayingP3c}. The algorithm solving (\ref{ProblemP3b}) with an optimization solver is referred to as EFA-S1, while the other (which solves the simplified (\ref{ProblemP3b}), i.e. the simplified source optimization) is referred to as EFA-S2. The function $C(\mathbf{\lambda}_f, \mathbf{\tilde{\lambda}}_{R,S})$ denotes the objective function (\ref{C_NonIF_Scalar_P2}). Since the optimization problems (\ref{ProblemP3a}), (\ref{ProblemP3b}) and the simplified source optimization are convex, $C(\mathbf{\lambda}_f, \mathbf{\tilde{\lambda}}_{R,S})$ monotonically decreases over iterations. Because (\ref{C_NonIF_Scalar_P2}) is lower-bounded, the two algorithms finally converge.

\section{Simplified NEFA Scheme}
\label{SecSimpRelayingWoEF}
This section proposes a simplified NEFA relaying (i.e. NEFA-S), considering uniform source power allocation. Similar to the simplified EFA schemes, the optimization problem is simplified to a power optimization by channel diagonalization. The original design problem of an NEFA scheme (i.e. problem (\ref{ProblemP1}), where {\small$\mathbf{Q}_D = \mathbf{0}$} and the inequality (\ref{FwdPowerConst_1P1}) is converted to equality) can be formulated as
{\small
\begin{IEEEeqnarray}{cl}
\label{OrigSimpRelayingWoEF}
\max_{\mathbf{Q}_S^{\prime}, \mathbf{F}^{\prime}}\,& \frac{1}{2}\log\det \Big(\mathbf{I} + (1\!-\!\rho)\mathbf{H}_{D, R} \mathbf{F}^{\prime} \mathbf{H}_{R, S} \mathbf{Q}_S^{\prime} \mathbf{H}_{R, S}^H \! \left[\mathbf{F}^{\prime}\right]^H \! \mathbf{H}_{D, R}^H \left[\mathbf{R}^{\prime}\right]^{-1} \Big)\IEEEyessubnumber\\
\text{s.t.} & \text{Tr}\left\{ (1 - \rho) \mathbf{F}^{\prime} \mathbf{H}_{R, S} \mathbf{Q}_S^{\prime} \mathbf{H}_{R, S}^H \left[\mathbf{F}^{\prime}\right]^H + \sigma_n^2\mathbf{F}^{\prime}\left[\mathbf{F}^{\prime}\right]^H \right\} = \nonumber \rho \text{Tr}\left\{\mathbf{H}_{R, S} \mathbf{Q}_S^{\prime} \mathbf{H}_{R, S}^H \right\}\,,\IEEEyessubnumber \label{FwdPowerConst_1P6}\\
&\text{Tr}\{\mathbf{Q}_S^{\prime}\} \leq P_S \,, \mathbf{Q}_S^{\prime} \succeq 0\,,\IEEEyessubnumber
\end{IEEEeqnarray}
}where {\small$\mathbf{R}^{\prime}\! =\! \sigma_{n}^2\mathbf{H}_{D, R} \mathbf{F}^{\prime} [\mathbf{F}^{\prime}]^H \mathbf{H}_{D, R}^H \! + \! \sigma_{n}^2\mathbf{I}$}, and {\small$\mathbf{F}^{\prime}$} denotes the relay processing matrix. Due to the absence of the energy flow {\small$\mathbf{Q}_D$}, the $S$-to-$R$ channel can be decomposed by SVD, such that {\small$\mathbf{H}_{R, S} \! = \! \mathbf{U}_{R,S} \mathbf{\Sigma}_{R,S} \mathbf{V}_{R,S}^H$}, where {\small$\mathbf{\Sigma}_{R,S} \! = \! diag\{\lambda_{R,S,1}, \ldots, \lambda_{R,S,r}\}$}. Take the EVD of {\small$\mathbf{H}_{R, S} \mathbf{Q}_S^{\prime} \mathbf{H}_{R, S}^H \! = \! \mathbf{\tilde{U}}_{R,S}^{\prime} \mathbf{\tilde{\Sigma}}_{R,S}^{\prime} [\mathbf{\tilde{U}}_{R,S}^{\prime}]^H$}, where {\small$\mathbf{\tilde{\Sigma}}_{R,S}^{\prime} \! = \! diag\{\tilde{\lambda}_{R,S,1}^{\prime}, \ldots, \tilde{\lambda}_{R,S,r}^{\prime}\}$}. It follows that {\small$\mathbf{Q}_S^{\prime} = \mathbf{V}_{R,S} \mathbf{\Sigma}_S^{\prime} \mathbf{V}_{R,S}^H$}, where {\small$\mathbf{\Sigma}_S^{\prime} = diag\{\lambda_{S,1}^{\prime}, \ldots, \lambda_{S,r}^{\prime}\}$}, {\small$\mathbf{\tilde{U}}_{R,S}^{\prime} \! = \! \mathbf{U}_{R,S}$}, and {\small$\mathbf{\tilde{\Sigma}}_{R,S}^{\prime} \! = \! \mathbf{\Sigma}_S^{\prime} \mathbf{\Sigma}_{R,S}$}. Due to the uniform source power allocation, {\small$\lambda_{S,m}^{\prime} \! = \! P_S/r$ $\forall m$}. Applying the above decomposition and omit the coefficient $1/2$, the power optimization of problem (\ref{OrigSimpRelayingWoEF}) at high receive SNR is formulated as
{\small
\begin{IEEEeqnarray}{ll}
\label{SimpRelayingWoEF}
\min_{\mathbf{\lambda}_f^{\prime}} \,&  -\!\sum_{m = 1}^{r}\!\log\!\left(\frac{(1\!-\!\rho) \tilde{\lambda}_{R,S,m}^{\prime} \lambda_{f,m}^{\prime} \lambda_{D,R,m}}{\sigma_n^2\left(1\!+\!\lambda_{f,m}^{\prime} \lambda_{D,R,m}\right)}\right) \IEEEyessubnumber\\
\text{s.t.} & \lambda_{f,1}^{\prime}, \lambda_{f,2}^{\prime}, \ldots, \lambda_{f,r}^{\prime} \geq 0\,, \IEEEyessubnumber\\
& \sum_{m = 1}^r\!\left(\!(1\!-\!\rho)\!\lambda_{f\!,m}^{\prime} \tilde{\lambda}_{R,S,m}^{\prime}\!+\!\sigma_n^2\lambda_{f\!,m}^{\prime}\right) \!=\!\sum_{m = 1}^r\! \rho\tilde{\lambda}_{R\!,S\!,m}^{\prime}\IEEEyessubnumber \label{EqRelayPwrConstP7a}
\end{IEEEeqnarray}
}where $\mathbf{\lambda}_f^{\prime} \! \triangleq \! [\lambda_{f,1}^{\prime}, \ldots, \lambda_{f,r}^{\prime}]^T$. The pairings of $\tilde{\lambda}_{R,S,m}^{\prime}$ and $\lambda_{D,R,m}$ for $m \!= \! 1,\ldots,r$ can be solved by Lemma \ref{Lemma1} with $P_D \! = \! 0$ and $\mathbf{\beta} \! = \! \mathbf{0}$. Hence, if $\tilde{\lambda}_{R,S,m}^{\prime} \! = \! P_S/r\cdot[\lambda_{R,S,\text{min}}, \ldots, \lambda_{R,S,\text{max}}]^T$ and $\lambda_{D,R,m}$ are arranged in an increasing order, the pairing problem is solved. Thus, the optimal $\lambda_{f,m}^{\prime}$ is obtained by
{\small
\begin{IEEEeqnarray}{l}
\label{EqClosedSoluP7a}
\left[\lambda_{f,m}^{\prime}\right]^{\star} = - \frac{1}{2\lambda_{D,R,m}} + \frac{1}{2}\sqrt{\frac{1}{\lambda^2_{D,R,m}}\!+\!\frac{4}{\left[\nu^{\prime}\right]^{\star}\lambda_{D,R,m}\left((1 - \rho)\tilde{\lambda}_{R,S,m}^{\prime}\!+\! \sigma_n^2\right)}}\,,
\end{IEEEeqnarray}
}where $\left[\nu^{\prime}\right]^{\star}$ satisfies (\ref{EqRelayPwrConstP7a}).

\section{Simulation Results}
\label{SecSimResults}
\begin{figure}[!t]
\centering
\subfigure[Achievable rate vs. $d_{DR}/d_{DS}$.]{
\label{Fig_RasFuncDR2DS_rRneqr_sym}
\includegraphics[width = 2.0in]
{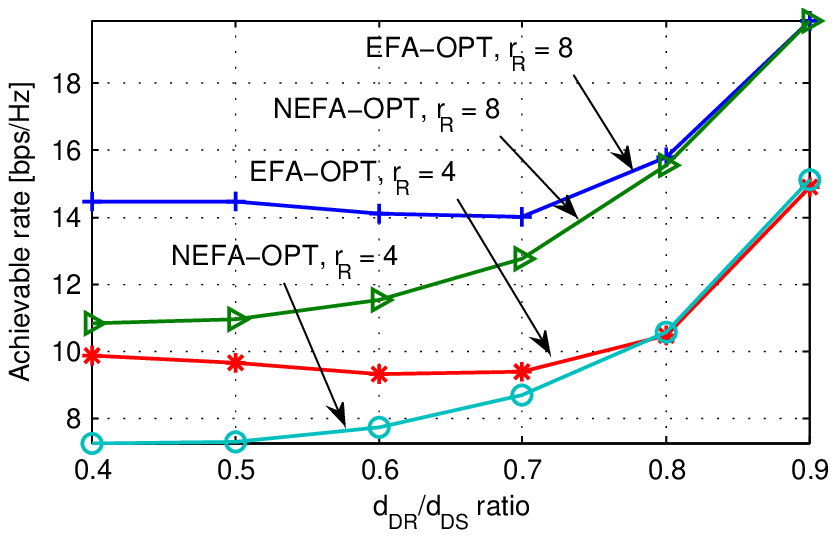}
}
\hfil
\subfigure[EFA-OPT: Percentage of the $\mathbf{Q}_S$ w/o eigenvalues (i.e. $\lambda_i(\mathbf{Q}_S)$) close to 0.]{
\label{Fig_FullRankPercentQs_DR2DS0p4n0p9}
\includegraphics[width = 2.0in]
{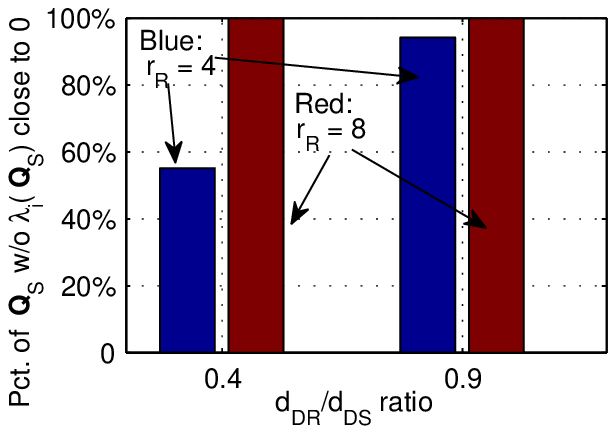}
}
\hfil
\subfigure[Best PS ratio vs. $d_{DR}/d_{DS}$.]{
\label{Fig_BestRhoAsFuncDR2DS_rRneqr_sym}
\includegraphics[width = 2.0in]
{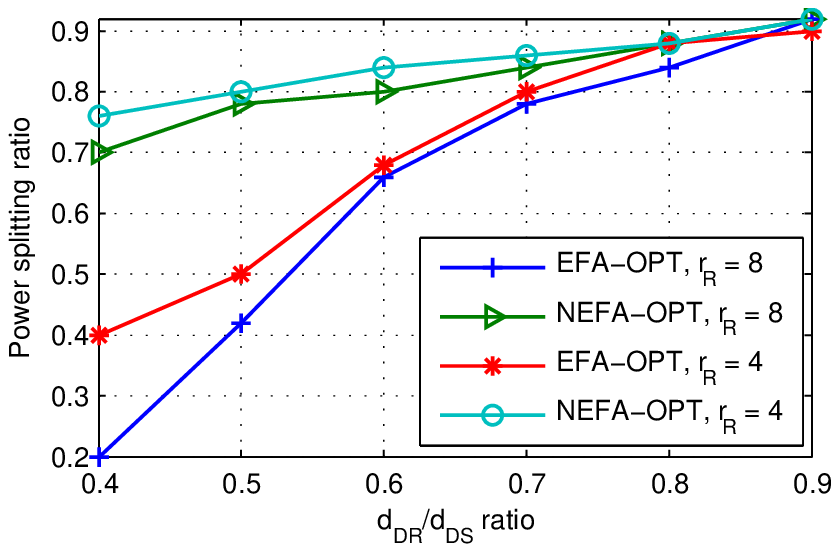}
}
\caption{Rate performance with different numbers of antennas at $R$ with $P_D = 0.5$\,W and $P_S = 0.1$\,W.}
\label{Fig_Performance_under_rR_sym}
\end{figure}

\begin{figure}[!t]
\centering
\subfigure[Rate of relay schemes when $r_R = 4$.]{
\label{Fig_RasFuncDR2DS_fair_rR4}
\includegraphics[width = 2.2in]
{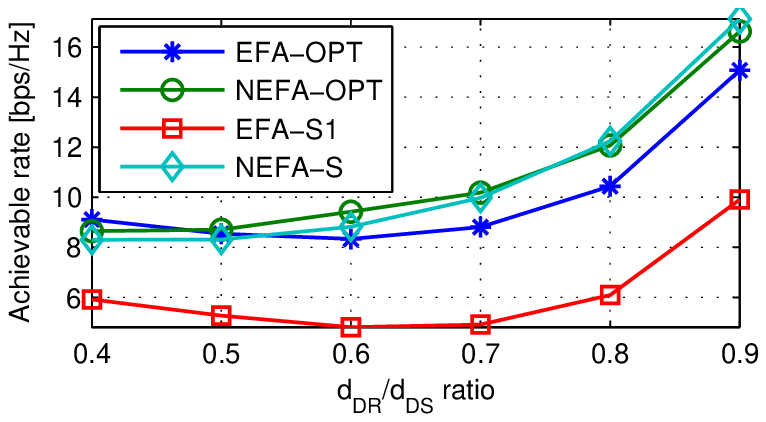}
}
\hfil
\subfigure[Effect of number of antennas at $R$.]{
\label{Fig_RasFuncDR2DS_fair_rR8}
\includegraphics[width = 2.2in]
{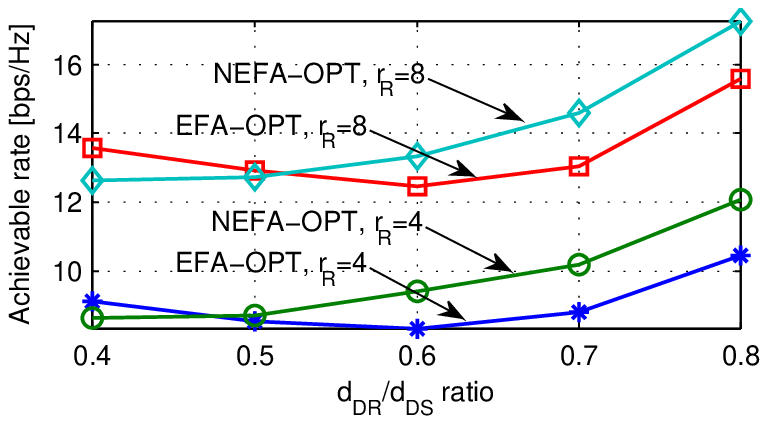}
}
\caption{Achievable Rate vs. $d_{DR}/d_{DS}$. For EFA schemes, $P_D = 0.1$\,W and $P_S = 0.1$\,W. For NEFA schemes, $P_S^{\prime} = 0.2$\,W.}
\label{Fig_Performance_under_AR2ABratio_fair}
\end{figure}

In the simulations, we assume broadside arrays are exploited, such that the channel matrix {\small$\mathbf{H}_{i,j} = \Lambda_{i,j}^{-1}(\sqrt{\frac{K}{1+K}}\mathbf{1} + \sqrt{\frac{1}{1+K}}\mathbf{\bar{H}}_{i,j})$}, where $\mathbf{1}$ (i.e. all-ones matrix) is the line-of-sight component, and {\small$\mathbf{\bar{H}}_{i,j}$} is the Rayleigh component. The large-scale fading is given by $\Lambda_{i,j}^{-1} = d_{ij}^{-3/2}$, where $d_{ij}$ is the distance between nodes $i$ and $j$ and $d_{DS} \triangleq d_{DR} + d_{RS}$.  The noise power $\sigma_n^2 = 1$\,$\mu$W, $d_{DS} = 10$\,m, the numbers of antennas at the terminals and $R$ are set as $r = r_R = 4$, the Rician factor $K=0$, unless otherwise stated. In the following Figs. \ref{Fig_Performance_under_AR2ABratio_sym}, \ref{Fig_Performance_under_AR2ABratio_asym}, \ref{Fig_Performance_under_AR2ABratio_fair} and  \ref{Fig_Performance_under_rR_sym}, the PS ratio $\rho$ is exhaustively searched among 0.02:0.02:0.98 to maximize the average rate.
%

\begin{figure*}[t]
\begin{minipage}[t][1.8in][t]{.49\textwidth}
\centering
\includegraphics[width = 2.0in]{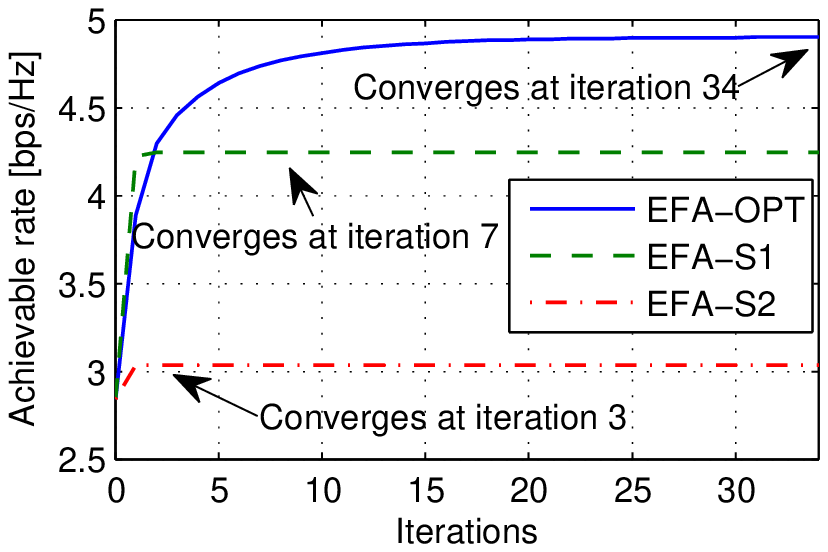}
\caption{Convergence example of EFA-OPT, EFA-S1, and EFA-S2.}
\label{Fig_Conv_Eg}
\end{minipage}\hfill
\begin{minipage}[t][1.8in][t]{.49\textwidth}
\centering
\includegraphics[width = 2.0in]{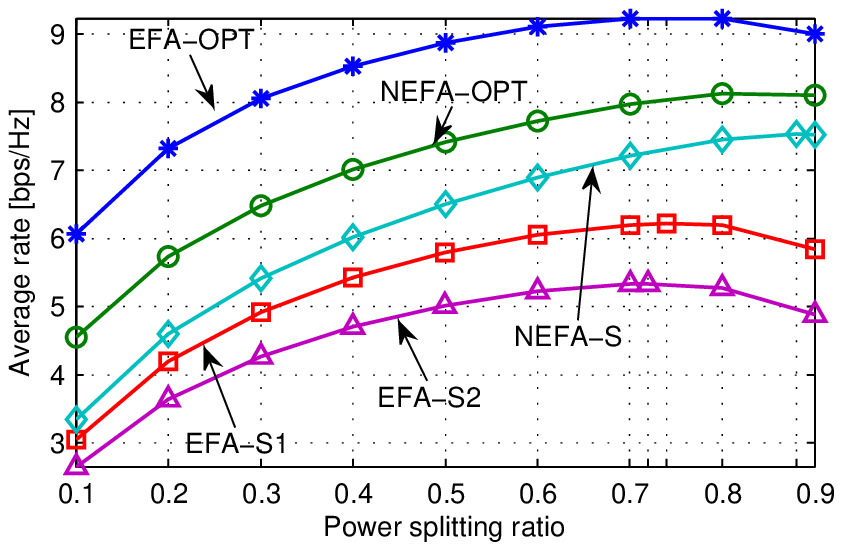}
\caption{Average rate as a function of PS ratio with $P_D = 0.5$\,W, $P_S = 0.1$\,W, and $d_{DR}/d_{DS} = 0.65$.}
\label{Fig_Rate_as_a_func_of_rho}
\end{minipage}
\end{figure*}
Fig. \ref{Fig_Conv_Eg} illustrates the convergence behaviors of EFA-OPT, EFA-S1, and EFA-S2, when $r=4$, $\text{SNR} = 20$\,dB ($d_{DR} = d_{RS} = 1$\,m, $P_S = P_D = 0.1$\,W, $\sigma_n^2 = 10^{-3}$\,W). Starting at the same initial point, EFA-OPT, EFA-S1, and EFA-S2 can converge after 34, 7, and 3 steps, respectively.

Fig. \ref{Fig_Rate_as_a_func_of_rho} investigates the average rate as a function of the PS ratio for a certain $d_{DR}/d_{DS}$. It is shown that the average rate curves of the five relaying schemes are concave over the PS ratios, reaching the maximum rates at PS ratios of 0.8, 0.8, 0.88, 0.74, and 0.72, respectively. The concave trend can be explained because a low PS ratio results in less available forwarding power, while a high PS ratio reduces the receive SNR at $R$. Both the above two factors can decrease the receive SNR at $D$.

\begin{figure*}[t]
\centering
\subfigure[Achievable rate vs. $d_{DR}/d_{DS}$, $K\!=\!0$.]{
\label{Fig_RasFuncDR2DS_sym}
\includegraphics[width = 2.0in]
{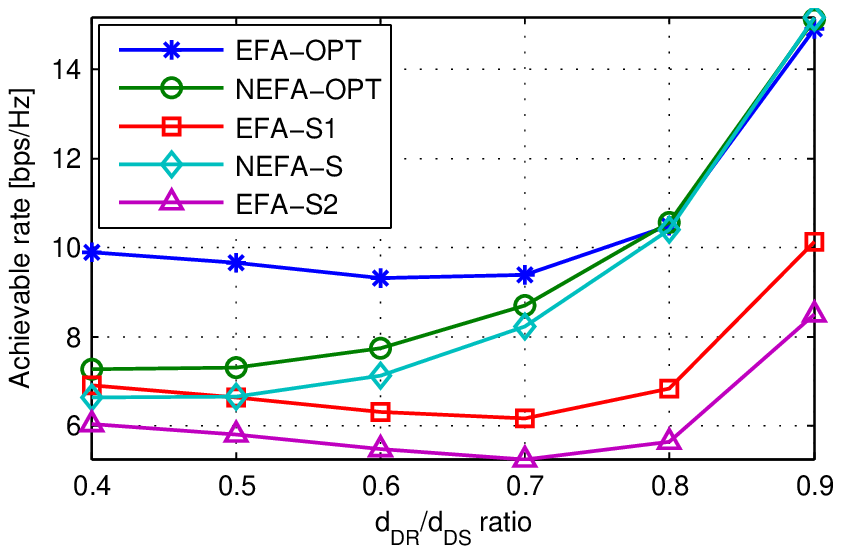}
}
\hfil
\subfigure[Best PS ratio vs. $d_{DR}/d_{DS}$, $K\!=\!0$.]{
\label{Fig_BestRhoAsFuncDR2DS_sym}
\includegraphics[width = 2.0in]
{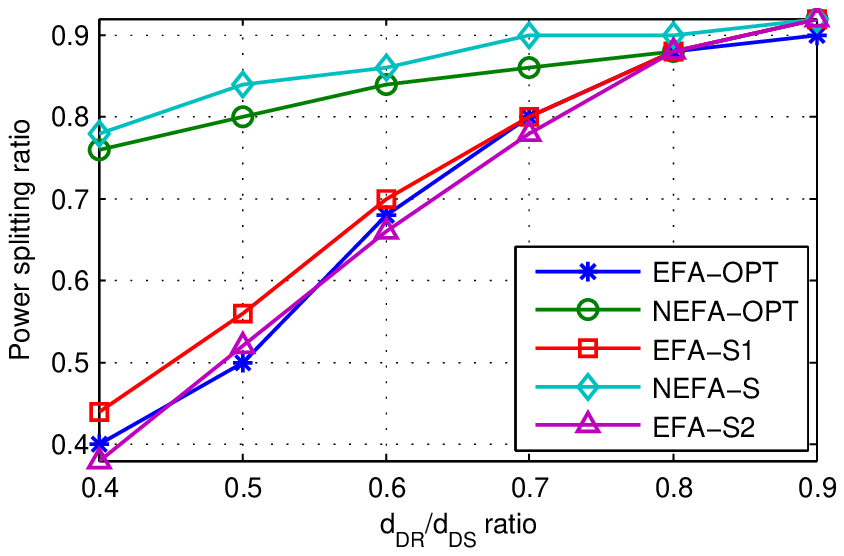}
}
\hfil
\subfigure[Achievable rate vs. $d_{\!D\!R}\!/\!d_{\!D\!S\!}$, $K\!=\!0.5$.]{
\label{Fig_RasFuncDR2DS_symK0p5}
\includegraphics[width = 2.0in]
{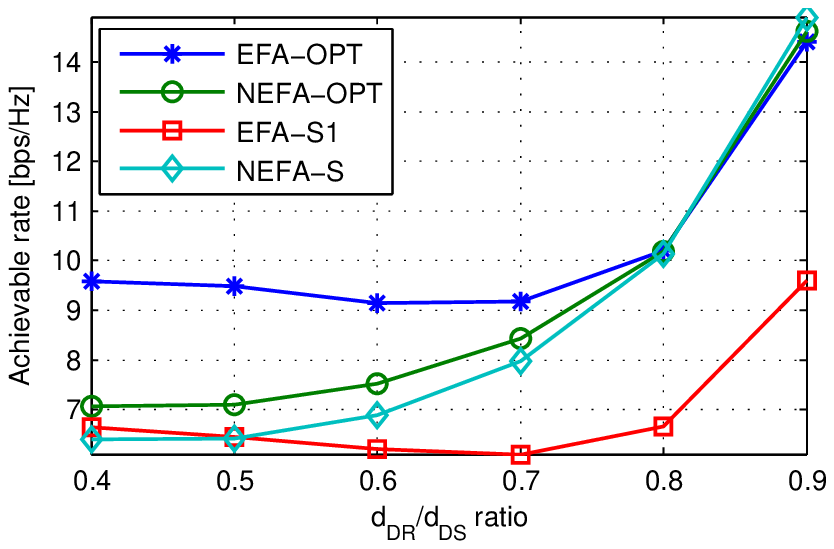}
}
\caption{Rate performance under different $d_{DR}/d_{DS}$ ratios with $P_D = 0.5$\,W and $P_S = 0.1$\,W.}
\label{Fig_Performance_under_AR2ABratio_sym}
\end{figure*}
Fig. \ref{Fig_RasFuncDR2DS_sym} shows the achievable rate as a function of $d_{DR}/d_{DS}$ ratio with symmetric power budgets at $D$ and $S$, when $K=0$. In general, the rates of the NEFA schemes (including NEFA-OPT and NEFA-S) decrease as $R$ moves towards $D$, because $R$ only extracts forwarding power from the information flow, and the reduced forwarding power degrades the rate. Different from the NEFA schemes, thanks to the EF, the rates of the EFA schemes (including EFA-OPT, EFA-S1, and EFA-S2) increase as $R$ moves towards $D$. When $R$ is close to $S$, similarly to the NEFA schemes, the rates of the EFA schemes can also increase as $d_{DR}/d_{DS}$ increases, because the EFA schemes can also harvest power from the source information flow. It is also observed in Fig. \ref{Fig_RasFuncDR2DS_sym} that the rate of EFA-OPT can be significantly higher than those of the NEFA schemes, when $d_{DR}/d_{DS} < 0.8$; while when $d_{DR}/d_{DS} \geq 0.8$, the rate of EFA-OPT is slightly lower than that of NEFA-OPT, the reason is discussed in the explanation of Fig. \ref{Fig_Performance_under_rR_sym}. The suboptimality of EFA-S1 and EFA-S2 makes the rate of EFA-S1 always lower than that of EFA-OPT and the rate of EFA-S2 always lower than that of EFA-S1. This is because in EFA-S1, the LSV of $\mathbf{\tilde{H}}_{R,S}$ is restricted to be $\mathbf{V}_{D,R}^{\ast}$, which reduces the searching area of the original feasible set of (\ref{FwdPowerConst_1P1}). Further, the source optimization (\ref{ProblemP3b}) of EFA-S1 requires the eigenvalues of $\mathbf{\tilde{H}}_{R,S}\mathbf{\tilde{H}}_{R,S}^H$ (recall that $\mathbf{\tilde{H}}_{R,S}$ is the $S$-to-$R$ effective channel) to be ordered in an increasing order, which tightens the constraint (\ref{EqldaRSineqConst}). These reasons lead to a reduced SNR at $R$. Thus, EFA-S1 is inferior to NEFA-OPT. As a further simplified version, the source power constraint (\ref{EqSrcPwrConst}) of EFA-S1 is further simplified to $\sum_{m = 1}^r \max\{|\mathbf{h}_{e,m}|^2\} \tilde{\lambda}_{R,S,m} \leq P_S$ in EFA-S2, which limits the received information signal power at $R$. The rate of NEFA-S is always lower than that of NEFA-OPT, because of the uniform source power allocation in NEFA-S. As $d_{DR}/d_{DS}$ increases, the gap between the rates of the NEFA schemes decreases. This is because with high-quality $S$-to-$R$ link, $S$ prefers to uniformly allocate the source power.

Fig. \ref{Fig_BestRhoAsFuncDR2DS_sym} demonstrates that when $R$ is close to $D$, the $S$-to-$R$ link is the critical link. To achieve high SNR at $D$, a low PS ratio should be selected. On the contrary, when $R$ is close to $S$, the $R$-to-$D$ link becomes the critical link, and the best PS ratios in this case are higher than those at low $d_{DR}/d_{DS}$ ratios. In general, the best PS ratios of the NEFA schemes are much higher than those of the EFA schemes at $d_{DR}/d_{DS} = 0.4$, because the NEFA schemes only harvests power from the information flow. When $R$ is close to $S$ (e.g. $d_{DR}/d_{DS}$ is $0.8$ or $0.9$), the gap between the best PS ratios of the EFA schemes and those of the NEFA schemes are negligible. This is because the EF heavily attenuates in these $d_{DR}/d_{DS}$ regions, both the EFA schemes and the NEFA schemes have to severely rely on the information flow for gaining the forwarding power. Fig. \ref{Fig_BestRhoAsFuncDR2DS_sym} also depicts that the best PS ratios of NEFA-S are higher than that of NEFA-OPT. This is because the uniform source power allocation makes NEFA-S suffer from a low-performance $S$-to-$R$ link. Thus, in order to improve the rate, the fraction of the signal power allocated to the EH receiver can be increased to enhance the forwarding power, such that the SNR at $D$ can be improved.

As shown in Fig. \ref{Fig_RasFuncDR2DS_symK0p5}, we also investigate the scenario where $K = 0.5$ \cite{TNB02}. The average achievable rates are similar to the scenario where $K = 0$. The best PS ratios in this scenario are also similar to Fig. \ref{Fig_BestRhoAsFuncDR2DS_sym}. Thus, the plot on PS ratio is omitted, due to the space constraint.

\begin{figure}[!t]
\centering
\subfigure[Achievable rate vs. $d_{DR}/d_{DS}$.]{
\label{Fig_RasFuncDR2DS_asym}
\includegraphics[width = 2.2in]
{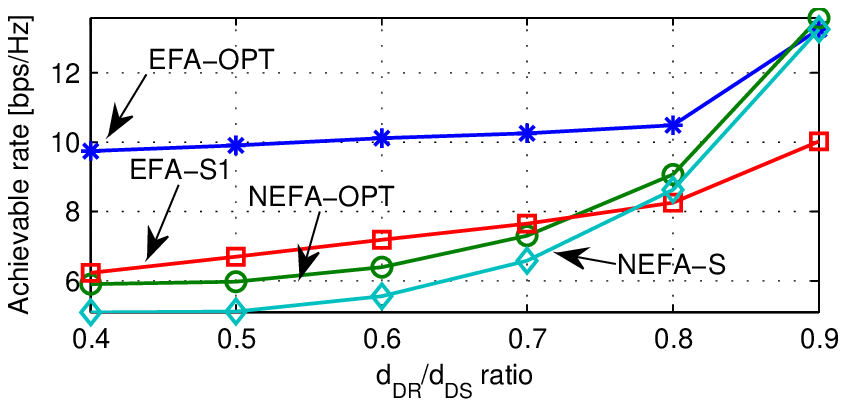}
}
\hfil
\subfigure[Best PS ratio vs. $d_{DR}/d_{DS}$.]{
\label{Fig_BestRhoAsFuncDR2DS_asym}
\includegraphics[width = 2.2in]
{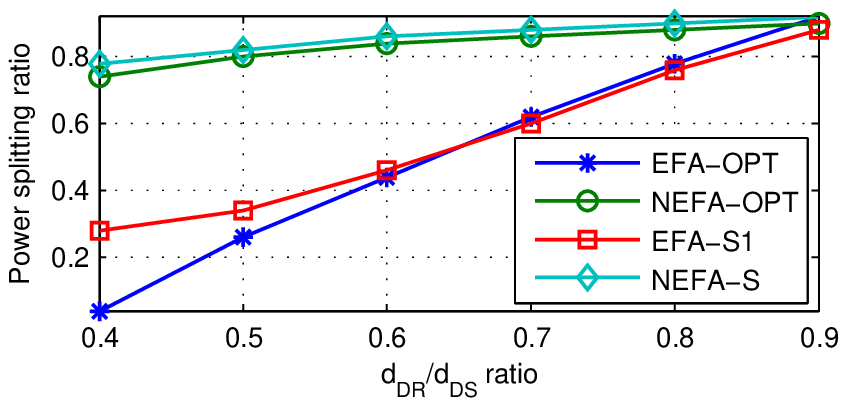}
}
\caption{Rate performance under different $d_{DR}/d_{DS}$ ratios with $P_D = 5$\,W and $P_S = 0.05$\,W.}
\label{Fig_Performance_under_AR2ABratio_asym}
\end{figure}
Fig. \ref{Fig_Performance_under_AR2ABratio_asym} studies the asymmetric power budgets scenario where $P_D$ is much greater than $P_S$. Intuitively, in such a scenario, the effect of the $S$-to-$R$ link on the end-to-end rate may not be as significant as in the symmetric case. Thus, as shown in Fig. \ref{Fig_RasFuncDR2DS_asym}, EFA-S1 can outperform the NEFA schemes rate-wise, when $d_{DR}/d_{DS} \leq 0.7$, although it suffers from the low $S$-to-$R$ link performance. It is also observed that the rate of EFA-OPT is higher than those of all the NEFA schemes, when $d_{DR}/d_{DS} \leq 0.8$. Different from the symmetric case, the rates of the EFA schemes always increase as $d_{DR}/d_{DS}$ increases, but not decrease and then increase as in Fig. \ref{Fig_RasFuncDR2DS_sym}. Because of the asymmetric power budgets, although the forwarding power decreases as $R$ moves towards $S$, the forwarding power can still make the rates scale with $d_{DR}/d_{DS}$. As shown in Fig. \ref{Fig_BestRhoAsFuncDR2DS_asym}, even when $d_{DR}/d_{DS} = 0.8$, the best PS ratios of the EFA schemes can be significantly lower than those of the NEFA schemes (which is different from Fig. \ref{Fig_BestRhoAsFuncDR2DS_sym}). This is because $P_D$ is much higher than $P_S$, and the power of the attenuated EF signal is still large enough to affect the forwarding power. It is shown that the gaps between the best PS ratio of the EFA schemes are significant at $d_{DR}/d_{DS}$ of $0.4$ and $0.5$. This is because EFA-OPT have a better $S$-to-$R$ link performance (i.e. the receive SNR at the ID receiver can be much higher), such that the rate can be improved by increasing the fraction of the signal power allocated to the ID receiver.

In the scenario where $r_R \geq r = 1$, our previous study \cite{HC15_CommLett} reveals that the EF is beneficial to the EFA scheme (i.e. the rate of the EFA scheme is significantly higher than that of the NEFA scheme) only when $r_R > 1$. In the MIMO relay system, although the EFA schemes can still benefit from the EF (i.e. the rates of EFA schemes can increase as $R$ moves towards $D$) when $r_R = r$, Fig. \ref{Fig_Performance_under_rR_sym} reveals that the presence of more antennas at $R$ (i.e. $r_R > r$) can further enhance the rate of the EFA scheme. It is observed in Fig. \ref{Fig_RasFuncDR2DS_rRneqr_sym} that when $d_{DR}/d_{DS} = 0.9$, the rate of EFA-OPT with $r_R = 4$ (i.e. $14.9028$) is slightly lower than that of NEFA-OPT with $r_R = 4$ (i.e. $15.1249$). A similar phenomenon can also be observed in Fig. \ref{Fig_RasFuncDR2DS_asym}. However, at $d_{DR}/d_{DS} = 0.9$, the rate of EFA-OPT with $r_R = 8$ (i.e. $19.8621$) is slightly higher than that of NEFA-OPT (i.e. $19.8408$). Analyzing Figs. \ref{Fig_RasFuncDR2DS_rRneqr_sym} and \ref{Fig_FullRankPercentQs_DR2DS0p4n0p9} reveals the reason. Fig. \ref{Fig_FullRankPercentQs_DR2DS0p4n0p9} studies the percentage of the $\mathbf{Q}_S$ (achieved by EFA-OPT) without eigenvalues (i.e. $\lambda_i(\mathbf{Q}_S)$ for $i = 1,\ldots, r$) close to 0 at $d_{DR}/d_{DS}$ ratios of $0.4$ and $0.9$. Such a $\mathbf{Q}_S$ without eigenvalues close to 0 indicates that no data stream is allocated with power close to 0. As shown in Fig. \ref{Fig_FullRankPercentQs_DR2DS0p4n0p9}, when $r_R = r$ and $d_{DR}/d_{DS} = 0.9$, in most cases, all the $r$ data streams at $S$ are allocated with considerably large power. Thus, in most cases, all the $r$ linearly combined data streams at $R$ are allocated with considerably high power. However, in EFA-OPT, except the $r$ data streams, there is one more EF signal being input into $R$. Since the dimension of the signal space at the ID receiver of $R$ is $r$, the EF leakage is totally combined with the $r$ data streams and amplified considerably. Although the retransmitted EF leakage can be canceled at $D$, it consumes lots of forwarding power. Nevertheless, when $r_R > r$, the dimension of the signal space at $R$ is $r_R \geq r + 1$, such that the EF leakage can be nearly aligned with a vector direction orthogonal to those of the linearly combined data streams. Thus, the EF leakage can be amplified with a smaller coefficient, and more power is consumed for the desired signal. Therefore, when $r_R = 8$, the rate of EFA-OPT is higher than that of NEFA-OPT. Despite the increased dimension of the signal space at $R$, the increase of $r_R$ also improves the information signal power, as well as the EF power, at $R$. Thus, as shown in Fig. \ref{Fig_RasFuncDR2DS_rRneqr_sym}, the rates of EFA-OPT and NEFA-OPT with $r_R = 8$ are higher than those of the schemes with $r_R = 4$, respectively. Fig. \ref{Fig_FullRankPercentQs_DR2DS0p4n0p9} also implies that an increase of $r_R$ enhances the $S$-to-$R$ link performance. Fig. \ref{Fig_BestRhoAsFuncDR2DS_rRneqr_sym} indicates that the best PS ratios of NEFA-OPT when $r_R = 8$ is lower than those with $r_R = 4$, due to the increased information signal power at $R$.

Fig. \ref{Fig_Performance_under_AR2ABratio_fair} studies the scenario where the EFA and the NEFA systems have the same total power budget. When the relay is close to $S$, the power of the harvested EF at $R$ is tiny in the EFA schemes, because of the \emph{high path loss}. The forwarding power at $R$ mainly comes from the source signal. Therefore, as shown in Fig. \ref{Fig_RasFuncDR2DS_fair_rR4}, the achievable rates of the EFA schemes are less than that of the NEFA schemes. When $R$ is close to $D$, the amount of the harvested EF power is high enough, such that the EFA-OPT can outperform NEFA-OPT rate-wise when $d_{DR}/d_{DS} = 0.4$. Fig. \ref{Fig_RasFuncDR2DS_fair_rR8} depicts that by increasing the number of antennas at $R$, harvested power at $R$ can be efficiently used to amplify the desired signal (as discussed in the explanation for Fig. \ref{Fig_Performance_under_rR_sym}), such that the rate difference between EFA-OPT and NEFA-OPT at $d_{DR}/d_{DS} = 0.4$ when $r_R = 8$ is larger than that when $r_R = 4$.

\section{Conclusion}
\label{SecConclusion}
In this paper, we have proposed the energy-flow-assisted (EFA) relaying protocol for the MIMO autonomous relay network, where the wireless-power relay node can relay the multiple source data streams and harvest the power for forwarding by processing the superposition of the energy flow (EF) from the destination and the source information signal. It is shown that contrary to the non-energy-flow-assisted (NEFA) relaying (where the relay only extracts power from the source signal for forwarding), the EF can significantly improve the rate of the EFA schemes, when the relay is close to the destination. It is also revealed that the additional antennas at the relay (i.e. number of antennas at the relay is greater than that at the terminals) can increase the dimension of the signal space at the information detecting receiver of the relay. By making use of the additional dimension, the information signal can be less interfered with the EF leakage, such that more power can be used to amplify and forward the desired information signal. Although the EFA scheme in this paper is studied from a communication theory and signal processing perspective and relies on several assumptions, the outcome of the research can be used as benchmarks for future studies, e.g. robust design for imperfect CSIT and practical impairments.

\appendix
\subsection{Proof of Theorem \ref{ConvergenceLimPointIterativeAlgorithm}}
\setcounter{equation}{0}
\renewcommand{\theequation}{A.\arabic{equation}}
\label{Appendix_ConvergenceLimPointIterativeAlgorithm}
Eq. (\ref{Eq_CloFormSolu_QCQPsubproF_constInactive}) implies that problem (\ref{QCQPsubproF_WMSEProBsF}) has infinite number of solutions. To achieve the unique solution $\mathbf{f}^\star$ to (\ref{QCQPsubproF_WMSEProBsF}), a tie-breaking rule can be included. Problem (\ref{SDRsubproBs_WMSEProBsF}) and the system of (\ref{EqSystem4Delta}) also have multiple solutions. Applying tie-breaking strategies, $\mathbf{X}_b^\star$ can be uniquely obtained by solving (\ref{SDRsubproBs_WMSEProBsF}), while the optimal rank-1 solution $\mathbf{X}_{b,0}$ can be uniquely derived from $\mathbf{X}_b^\star$ by the rank reduction. Hence, the global optimal solution $\mathbf{b}^\star$ to (\ref{QCQPsubproBs_WMSEProBsF}) is uniquely attained.

Due to the compactness of $\mathbf{x}$, there exists a limit point {\small$\mathbf{\bar{x}} = (\mathbf{\bar{x}}_1, \mathbf{\bar{x}}_2, \mathbf{\bar{x}}_3, \mathbf{\bar{x}}_4)$} such that $\mathbf{x}^{(\kappa)}$ converges to $\mathbf{\bar{x}}$ as $\kappa$ tends to infinity (i.e. {\small$\mathbf{x}^{(\kappa)} \rightarrow \mathbf{\bar{x}}$}). Because of the convergence shown in Theorem \ref{ConvergenceIterativeAlgorithm}, we have {\small$C_{\text{iter}}(\mathbf{x}^{(\kappa)}) \rightarrow C_{\text{iter}}(\mathbf{\bar{x}})$}.
Proving the convergence of {\small$\{\mathbf{x}^{(\kappa)}\}_{\kappa = 0}^\infty$} is to show that if {\small$\mathbf{x}^{(\kappa)} \rightarrow \mathbf{\bar{x}}$}, {\small$\mathbf{x}^{(\kappa + 1)} \rightarrow \mathbf{\bar{x}}$}. Due to (\ref{EqOptimA_0}) and (\ref{EqOptimW}), problems (\ref{SubproA_0_WMSEProBsF}) and (\ref{SubproW_WMSEProBsF}) have unique solutions. Thanks to the tie-breaking rule, problem (\ref{QCQPsubproF_WMSEProBsF}) also has an unique optimal solution. Thus, by using the contradiction method in \cite{Bertsekas99}, it can be easily shown that if {\small$\mathbf{x}^{(\kappa)} \rightarrow \mathbf{\bar{x}}$}, {\small$\mathbf{y}_1^{(\kappa + 1)} \rightarrow \mathbf{\bar{x}}$}; if {\small$\mathbf{y}_1^{(\kappa + 1)} \rightarrow \mathbf{\bar{x}}$}, {\small$\mathbf{y}_2^{(\kappa + 1)} \rightarrow \mathbf{\bar{x}}$}; if {\small$\mathbf{y}_2^{(\kappa + 1)} \rightarrow \mathbf{\bar{x}}$}, {\small$\mathbf{y}_3^{(\kappa + 1)} \rightarrow \mathbf{\bar{x}}$}.
Then, it remains to show that if {\small$\mathbf{y}_3^{(\kappa + 1)} \rightarrow \mathbf{\bar{x}}$}, {\small$\mathbf{x}^{(\kappa + 1)} \rightarrow \mathbf{\bar{x}}$}. Recall that when solving the subproblem of $\mathbf{B}_S$, {\small$\mathbf{x}_4^{(\kappa)}$} and {\small$\mathbf{x}_4^{(\kappa + 1)}$} are extracted from the optimal rank-1 matrices {\small$\mathbf{X}_{b,0}^{(\kappa)}$} and {\small$\mathbf{X}_{b,0}^{(\kappa + 1)}$}, respectively. Let {\small$\mathbf{y}_{B, 3}^{(\kappa + 1)} \triangleq (\mathbf{x}_1^{(\kappa + 1)}, \mathbf{x}_2^{(\kappa + 1)}, \mathbf{x}_3^{(\kappa + 1)}, \text{vec}(\mathbf{X}_{b,0}^{(\kappa)})^T)$}, {\small$\mathbf{y}_{B, 4}^{(\kappa + 1)} \triangleq (\mathbf{x}_1^{(\kappa + 1)}, \mathbf{x}_2^{(\kappa + 1)}, \mathbf{x}_3^{(\kappa + 1)}, \text{vec}(\mathbf{X}_{b,0}^{(\kappa + 1)})^T)$} and {\small$\mathbf{\bar{X}}_{b,0} \! = \! [\mathbf{\bar{x}}_4^T\mathbf{\bar{x}}_4^\ast, \mathbf{\bar{x}}_4^T; \mathbf{\bar{x}}_4^\ast, 1]$}. Proving the above claim ends up showing that if {\small$\mathbf{y}_{B, 3}^{(\kappa + 1)} \rightarrow (\mathbf{\bar{x}}_1, \mathbf{\bar{x}}_2, \mathbf{\bar{x}}_3, \text{vec}(\mathbf{\bar{X}}_{b,0})^T)$}, {\small$\mathbf{y}_{B, 4}^{(\kappa + 1)} \rightarrow (\mathbf{\bar{x}}_1, \mathbf{\bar{x}}_2, \mathbf{\bar{x}}_3, \text{vec}(\mathbf{\bar{X}}_{b,0})^T)$}.
This is then proved by contradiction. Assuming that the above claim is not true, there always exists a non-zero scalar $e_0$ such that {\small$\|\mathbf{X}_{b,0}^{(\kappa + 1)} - \mathbf{X}_{b,0}^{(\kappa)}\|_F \geq e_0$}. Let {\small$\mathbf{Z} = (\mathbf{X}_{b,0}^{(\kappa + 1)} - \mathbf{X}_{b,0}^{(\kappa)})/ \|\mathbf{X}_{b,0}^{(\kappa + 1)} - \mathbf{X}_{b,0}^{(\kappa)}\|_F$} such that {\small$\mathbf{Z} \rightarrow \mathbf{\bar{Z}}$}. By fixing a $\theta \in [0, e_0]$, we can obtain a point {\small$\mathbf{X}_{b,0}^{(\kappa)} + \theta \mathbf{Z}$} lying in the segment of {\small$\mathbf{X}_{b,0}^{(\kappa + 1)}$} and {\small$\mathbf{X}_{b,0}^{(\kappa)}$}. Since the feasible set of problem (\ref{SDRsubproBs_WMSEProBsF}) is convex, the point {\small$\mathbf{X}_{b,0}^{(\kappa)} + \theta \mathbf{Z}$} is within this feasible set. Denote the objective function of (\ref{SDRsubproBs_WMSEProBsF}) as {\small$C_B(\mathbf{X}_b; \mathbf{x}_1, \mathbf{x}_2, \mathbf{x}_3)$}. Since $(\mathbf{x}_1, \! \mathbf{x}_2, \! \mathbf{x}_3)$ in {\small$\mathbf{y}_{B, 3}^{(\kappa + 1)}$ and $\mathbf{y}_{B, 4}^{(\kappa + 1)}$} are fixed as {\small$(\mathbf{x}_1^{(\kappa + 1)}, \! \mathbf{x}_2^{(\kappa + 1)}, \! \mathbf{x}_3^{(\kappa + 1)})$}, the notation of this objective function is simplified as {\small$C_B(\mathbf{X}_b)$}. Due to the optimality of the rank-1 {\small$\mathbf{X}_{b,0}^{(\kappa + 1)}$}, we have {\small$C_B(\mathbf{X}_{b,0}^{(\kappa + 1)})\!\leq\!C_B(\mathbf{X}_{b,0}^{(\kappa)} \! + \! \theta \mathbf{Z})$}. Meanwhile, because {\small$C_B(\mathbf{X}_b)$} is convex, {\small$C_B(\mathbf{X}_{b,0}^{(\kappa)} \! + \! \theta \mathbf{Z}) \! \leq \! C_B(\mathbf{X}_{b,0}^{(\kappa)})$} \cite{BV04}. Thus,
{\small
\begin{equation}
\label{EqInequalitiesSegment}
C_B(\mathbf{X}_{b,0}^{(\kappa + 1)}) \leq C_B(\mathbf{X}_{b,0}^{(\kappa)} + \theta \mathbf{Z}) \leq C_B(\mathbf{X}_{b,0}^{(\kappa)})\,.
\end{equation}
}Because {\small$C_{\text{iter}}(\mathbf{x}^{(\kappa)}) \rightarrow C_{\text{iter}}(\mathbf{\bar{x}})$} and (\ref{EqC_iterMonoDecrease}), {\small$C_{\text{iter}}(\mathbf{y}_{B, 3}^{(\kappa + 1)}) = C_B(\mathbf{X}_{b,0}^{(\kappa)}) + C_0(\mathbf{x}_1^{(\kappa + 1)}, \mathbf{x}_2^{(\kappa + 1)}, \mathbf{x}_3^{(\kappa + 1)}) \rightarrow C_{\text{iter}}(\mathbf{\bar{x}}) = C_B(\mathbf{\bar{X}}_{b,0}) + C_0(\mathbf{\bar{x}}_1, \mathbf{\bar{x}}_2, \mathbf{\bar{x}}_3)$}, where {\small$C_0(\cdot)$} denotes the other terms not containing $\mathbf{\bar{X}}_{b,0}$. Hence, {\small$C_B(\mathbf{X}_{b,0}^{(\kappa)}) \rightarrow C_B(\mathbf{\bar{X}}_{b,0})$}. Because of (\ref{EqInequalitiesSegment}), the value of {\small$C_B(\mathbf{X}_{b,0}^{(\kappa + 1)})$} also converges to {\small$C_B(\mathbf{\bar{X}}_{b,0})$}. Taking the limit of (\ref{EqInequalitiesSegment}) as $\kappa$ tends to infinity yields {\small$C_B(\mathbf{\bar{X}}_{b,0}) \leq C_B(\mathbf{\bar{X}}_{b,0} + \theta \mathbf{\bar{Z}}) \leq C_B(\mathbf{\bar{X}}_{b,0})$}, i.e. {\small$C_B(\mathbf{\bar{X}}_{b,0} + \theta \mathbf{\bar{Z}}) = C_B(\mathbf{\bar{X}}_{b,0})$}. This means that given {\small$\text{vec}(\mathbf{A}_0)^T = \mathbf{\bar{x}}_1$}, {\small$\text{vec}(\mathbf{W})^T = \mathbf{\bar{x}}_2$}, {\small$\text{vec}(\mathbf{F})^T = \mathbf{\bar{x}}_3$}, both the high-rank {\small$\mathbf{\bar{X}}_{b,0} + \theta \mathbf{\bar{Z}}$} and the rank-1 {\small$\mathbf{\bar{X}}_{b,0}$} are the optimal solutions of problem (\ref{SDRsubproBs_WMSEProBsF}).
Next, making use of contradiction, we show that the optimal rank-1 solution derived from {\small$\mathbf{\bar{X}}_{b,0} + \theta \mathbf{\bar{Z}}$} is different from {\small$\mathbf{\bar{X}}_{b,0}$}. Thus, assume the contrary, i.e. {\small$\mathbf{\bar{X}}_{b,0}$} is the rank-1 solution derived from {\small$\mathbf{\bar{X}}_{b,0} + \theta \mathbf{\bar{Z}}$}. Since the rank update rules of (\ref{EqRankReductionUpdate}) and (\ref{EqSystem4Delta}) preserve the primal feasibility (i.e. {\small$\text{Tr}\{\mathbf{B}_m \mathbf{X}_b\} = \text{Tr}\{\mathbf{B}_m \mathbf{X}_{b,0}\}$} for $m = 2,3,4$)\cite{HP10}, it follows that {\small$\text{Tr}\{\mathbf{B}_m (\mathbf{\bar{X}}_{b,0} + \theta \mathbf{\bar{Z}})\} = \text{Tr}\{\mathbf{B}_m \mathbf{\bar{X}}_{b,0}\}$}, namely, {\small$\text{Tr}\{\mathbf{B}_m \mathbf{\bar{Z}}\} = 0$}. Recall that we have assumed that {\small$\mathbf{X}_{b,0}^{(\kappa)}$} and {\small$\mathbf{X}_{b,0}^{(\kappa + 1)}$} converge to different limit points. Let {\small$\mathbf{X}_{b,0}^{(\kappa + 1)}$} converge to another rank-1 matrix {\small$\mathbf{\bar{X}}_{b,0}^\prime$}. It follows that {\small$\text{Tr}\{\mathbf{B}_m \mathbf{\bar{X}}_{b,0}^\prime\} = \text{Tr}\{\mathbf{B}_m \mathbf{\bar{X}}_{b,0}\}$}, which implies that two rank-1 solutions can be derived from one high-rank optimal solution by the rank reduction procedure. This contradicts to the hypothesis that (with a tie-breaking strategy) the rank reduction procedure yields an unique rank-1 solution. Thus, the rank-1 solution derived from {\small$\mathbf{\bar{X}}_{b,0} + \theta \mathbf{\bar{Z}}$} is different from {\small$\mathbf{\bar{X}}_{b,0}$}. However, this claim contradicts to the hypothesis that solving problem (\ref{SDRsubproBs_WMSEProBsF}) only yield an unique rank-1 solution. This contradiction illustrates that if {\small$\mathbf{y}_{B, 3}^{(\kappa + 1)} \rightarrow (\mathbf{\bar{x}}_1, \mathbf{\bar{x}}_2, \mathbf{\bar{x}}_3, \text{vec}(\mathbf{\bar{X}}_{b,0})^T)$}, {\small$\mathbf{y}_{B, 4}^{(\kappa + 1)} \rightarrow (\mathbf{\bar{x}}_1, \mathbf{\bar{x}}_2, \mathbf{\bar{x}}_3, \text{vec}(\mathbf{\bar{X}}_{b,0})^T)$}. Consequently, we conclude that if {\small$\mathbf{x}^{(\kappa)} \rightarrow \mathbf{\bar{x}}$}, {\small$\mathbf{x}^{(\kappa + 1)} \rightarrow \mathbf{\bar{x}}$}.

\subsection{Proof of Theorem \ref{LimUnNCEstationary}}
\label{Appendix_LimUnNCEstationary}
\setcounter{equation}{0}
\renewcommand{\theequation}{B.\arabic{equation}}
Let {\small$\mathbf{A}_0\!= \! \Psi_{\mathbf{A}_0}(\mathbf{W}, \mathbf{F}, \mathbf{B}_S)$}, {\small$\mathbf{W} \! = \! \Psi_{\mathbf{W}}(\mathbf{A}_0, \mathbf{F}, \mathbf{B}_S)$}, {\small$\mathbf{F} \! = \! \Psi_{\mathbf{F}}(\mathbf{A}_0, \mathbf{W}, \mathbf{B}_S)$}, and {\small$\mathbf{B}_S \! = \! \Psi_{\mathbf{B}_S}(\mathbf{A}_0, \mathbf{W}, \mathbf{F})$} represent subproblems (\ref{SubproA_0_WMSEProBsF}), (\ref{SubproW_WMSEProBsF}), (\ref{OrigSubproF_WMSEProBsF}), and (\ref{OrigSubproBs_WMSEProBsF}), respectively. It is shown in Appendix \ref{Appendix_ConvergenceLimPointIterativeAlgorithm} that {\small$\mathbf{y}_1^{(\kappa + 1)}$}, {\small$\mathbf{y}_2^{(\kappa + 1)}$}, {\small$\mathbf{y}_3^{(\kappa + 1)}$} and {\small$\mathbf{x}^{(\kappa + 1)}$} converge to $\mathbf{\bar{x}}$. Hence, we have {\small$\mathbf{\bar{A}}_0 = \Psi_{\mathbf{A}_0}(\mathbf{\bar{W}}, \mathbf{\bar{F}}, \mathbf{\bar{B}_S})$}, {\small$\mathbf{\bar{W}} = \Psi_{\mathbf{\bar{A}}_0, \mathbf{W}}(\mathbf{\bar{F}}, \mathbf{\bar{B}}_S)$}, {\small$\mathbf{\bar{F}} = \Psi_{\mathbf{F}}(\mathbf{\bar{A}}_0, \mathbf{\bar{W}}, \mathbf{\bar{B}}_S)$}, and {\small$\mathbf{\bar{B}}_S = \Psi_{\mathbf{B}_S}(\mathbf{\bar{A}}_0, \mathbf{\bar{W}}, \mathbf{\bar{F}})$}. Thus, in the subproblems of $\mathbf{A}_0$ and $\mathbf{W}$, $\mathbf{\bar{A}}_0$ and $\mathbf{\bar{W}}$ respectively satisfy corresponding Karush-Kuhn-Tucker (KKT) conditions such that
{\small$\nabla_{\mathbf{A}_0} (\text{Tr}\{\mathbf{\bar{A}}_0 \mathbf{E}(\mathbf{\bar{W}}, \mathbf{\bar{F}}, \mathbf{\bar{B}}_S)\} - \log\det(\mathbf{\bar{A}}_0)) = 0$} and {\small$\nabla_{\mathbf{W}} (\text{Tr}\{\mathbf{\bar{A}}_0 \mathbf{E}(\mathbf{\bar{W}}, \mathbf{\bar{F}}, \mathbf{\bar{B}}_S)\}) = 0$}.
Let {\small$g_R(\mathbf{F}, \mathbf{B}_S) \triangleq \text{Tr}\{ (1 - \rho)\mathbf{F}\mathbf{H}_{R, S} \mathbf{B}_S \mathbf{B}_S^H \cdot \mathbf{H}_{R, S}^H \mathbf{F}^H + (1 - \rho) \mathbf{F} \mathbf{H}_{R, D} \mathbf{Q}_D \mathbf{H}_{R, D}^H \mathbf{F}^H + \sigma_n^2\mathbf{F}\mathbf{F}^H \} - \rho \text{Tr}\{\mathbf{H}_{R, D} \mathbf{Q}_D \mathbf{H}_{R, D}^H + \mathbf{H}_{R, S}\mathbf{B}_S \mathbf{B}_S^H \mathbf{H}_{R, S}^H \}$} and {\small$g_S(\mathbf{B}_S) \triangleq \text{Tr}\{\mathbf{B}_S\mathbf{B}_S^H\} - P_S$}. The Lagrangian of problem (\ref{OrigSubproF_WMSEProBsF}) is given by {\small$L_{\mathbf{F}}(\mathbf{F}, \xi_1) = \text{Tr}\{\mathbf{\bar{A}}_0 \mathbf{E}(\mathbf{\bar{W}}, \mathbf{F},$ $\mathbf{\bar{B}}_S)\} + \xi_1 g_R(\mathbf{F}, \mathbf{\bar{B}}_S)$}. Thus, $\mathbf{\bar{F}}$ and the associated optimal Lagrangian multiplier $\bar{\xi}_1$ must satisfy the KKT conditions given by
{\small$\nabla_{\mathbf{F}^\ast}(\text{Tr}\{\mathbf{\bar{A}}_0 \mathbf{E}(\mathbf{\bar{W}}, \mathbf{\bar{F}}, \mathbf{\bar{B}}_S)\}) + \bar{\xi}_1 \nabla_{\mathbf{F}^\ast} g_R(\mathbf{\bar{F}}, \mathbf{\bar{B}}_S) = 0$} and {\small$\bar{\xi}_1 \geq 0\,,\, g_R(\mathbf{\bar{F}}, \mathbf{\bar{B}}_S) \leq 0\,,\, \bar{\xi}_1 g_R(\mathbf{\bar{F}}, \mathbf{\bar{B}}_S) = 0$}.
The Lagrangian of problem (\ref{OrigSubproBs_WMSEProBsF}) is: {\small$L_{\mathbf{B}_S}(\mathbf{B}_S, \xi_0, \epsilon_0) = \text{Tr}\{\mathbf{\bar{A}}_0 \mathbf{E}(\mathbf{\bar{W}}, \mathbf{\bar{F}}, \mathbf{\bar{B}}_S)\} + \xi_2 g_R(\mathbf{\bar{F}}, \mathbf{B}_S) + \epsilon_2 g_S(\mathbf{B}_S)$}. Hence, $\mathbf{\bar{B}}_S$ and the associated optimal multipliers $\bar{\xi}_2$ and $\bar{\epsilon}_2$ must satisfy the KKT conditions listed as follows.
{\small$\nabla_{\mathbf{B}_S^\ast}(\text{Tr}\{\mathbf{\bar{A}}_0 \mathbf{E}(\mathbf{\bar{W}}, \mathbf{\bar{F}}, \mathbf{\bar{B}}_S)\}) + \bar{\xi}_2 \nabla_{\mathbf{B}_S^\ast} g_R(\mathbf{\bar{F}}, \mathbf{\bar{B}}_S) + \bar{\epsilon_2} \nabla_{\mathbf{B}_S^\ast} g_S(\mathbf{\bar{B}}_S) = 0$},
{\small$\bar{\xi}_2 \geq 0\,,\, g_R(\mathbf{\bar{F}}, \mathbf{\bar{B}}_S) \leq 0\,,\, \bar{\xi}_2 g_R(\mathbf{\bar{F}}, \mathbf{\bar{B}}_S) = 0$}, and
{\small$\bar{\epsilon}_2 \geq 0\,,\, g_S(\mathbf{\bar{B}}_S) \leq 0\,,\, \bar{\epsilon}_2 g_S(\mathbf{B}_S) = 0$}.
The complementary slackness conditions in the KKT conditions of {\small$\mathbf{\bar{F}} = \Psi_{\mathbf{F}}(\mathbf{\bar{A}}_0, \mathbf{\bar{W}}, \mathbf{\bar{B}}_S)$} and {\small$\mathbf{\bar{B}}_S = \Psi_{\mathbf{B}_S}(\mathbf{\bar{A}}_0, \mathbf{\bar{W}}, \mathbf{\bar{F}})$} implies that when {\small$g_R(\mathbf{\bar{F}}, \mathbf{\bar{B}}_S) < 0$}, {\small$\nabla_{\mathbf{F}^\ast} g_R(\mathbf{\bar{F}}, \mathbf{\bar{B}}_S)$} and {\small$\nabla_{\mathbf{B}_S^\ast} g_R(\mathbf{\bar{F}}, \mathbf{\bar{B}}_S)$} are inactive in the Lagrangian functions for $\mathbf{F}$ and $\mathbf{B}_S$. Thus, under this condition, combining the KKT conditions of {\small$\mathbf{\bar{A}}_0 = \Psi_{\mathbf{A}_0}(\mathbf{\bar{W}}, \mathbf{\bar{F}}, \mathbf{\bar{B}_S})$}, {\small$\mathbf{\bar{W}} = \Psi_{\mathbf{\bar{A}}_0, \mathbf{W}}(\mathbf{\bar{F}}, \mathbf{\bar{B}}_S)$}, {\small$\mathbf{\bar{F}} = \Psi_{\mathbf{F}}(\mathbf{\bar{A}}_0, \mathbf{\bar{W}}, \mathbf{\bar{B}}_S)$} and {\small$\mathbf{\bar{B}}_S = \Psi_{\mathbf{B}_S}(\mathbf{\bar{A}}_0, \mathbf{\bar{W}}, \mathbf{\bar{F}})$} shows that {\small$(\mathbf{\bar{A}}_0, \mathbf{\bar{W}}, \mathbf{\bar{F}}, \mathbf{\bar{B}}_S, \bar{\epsilon}_2)$} satisfies the KKT conditions of problem (\ref{WMSEProBsF}).

\subsection{Proof of Lemma \ref{Lemma1}}
\label{Appendix_ProofLemma1}
\setcounter{equation}{0}
\renewcommand{\theequation}{C.\arabic{equation}}
In the subsequent part, it is defined that {\small$a_m \! \triangleq \! \lambda_{D,R,m}$} and {\small$h(z_m, \nu, a_m) \! \triangleq \! a_m \lambda^{\star}_f(z_m, \nu, a_m)/(1 + \lambda^{\star}_f(z_m,$} {\small$\nu, a_m) a_m)$}, where {\small$\lambda^{\star}_f(\cdot)$} denotes (\ref{EqClosedSoluP3a}). Since the objective function of (\ref{ProblemP3a}) equals {\small$\log[ \left((1\!-\!\rho)/\sigma_n^2\right)^r\cdot$ $\prod(\tilde{\lambda}_{R,S,m} h(z_m, \nu, a_m))]$} and $\log(\cdot)$ monotonically increases, proving Lemma \ref{Lemma1} ends up showing
{\small\begin{equation}
\label{EqEqualConstLemma1}
h(z_i, \nu_1, a_p)h(z_j, \nu_1, a_q) \prod \left[h(z_m, \nu_1, a_n)\right]\geq h(z_j, \nu_2, a_p)h(z_i, \nu_2, a_q)\prod \left[h(z_m, \nu_2, a_n)\right]\,,
\end{equation}
}where $m \neq i,j$, $n \neq p,q$; $\nu_1$ and $\nu_2$ optimal multipliers corresponding to $\pi_1(\mathbf{z})$ and $\pi_2(\mathbf{z})$. Since the power allocation at the source is fixed, the r.h.s. of the equality constraint (\ref{EqEqConstP3a}) equals a constant. Thus, the l.h.s. of (\ref{EqEqConstP3a}) (which is a function of $\nu$ and $\pi(\mathbf{z})$) with $\nu_1$ and $\pi_1(\mathbf{z})$ is equal to that with $\nu_2$ and $\pi_2(\mathbf{z})$. That is, $\nu_1$ and $\nu_2$ conform to {\small$z_i\lambda^{\star}_f(z_i, \nu_1, a_p) + z_j\lambda^{\star}_f(z_j, \nu_1, a_q) - z_j\lambda^{\star}_f(z_j, \nu_2, a_p) - z_i\lambda^{\star}_f(z_i, \nu_2, a_q) = \sum [\sqrt{z_m} (\sqrt{\nu_1\nu_2 z_m + 4a_n\nu_1} - \sqrt{\nu_1\nu_2 z_m + 4a_n\nu_2})/(2a_n\sqrt{\nu_1\nu_2})]$}, where $m \neq i,j$ and $n \neq p,q$. The above equality reveals a constraint on $\nu_1$ and $\nu_2$: if $\nu_1 \leq \nu_2$, the l.h.s. of the above equality is no greater than 0; otherwise, its l.h.s. is no less than 0.

When $0 < \nu_1 \leq \nu_2$, $\partial h(z_m, \nu, a_n)/ \partial\nu < 0$. Therefore, the proof of Lemma \ref{Lemma1} ends up showing $h(z_i, \nu_1, a_p)h(z_j, \nu_1, a_q) \geq h(z_j, \nu_2, a_p)h(z_i, \nu_2, a_q)$. After manipulation, proving the above inequality becomes to show {\small$\nu_2\sqrt{\nu_2 z_i + 4a_q}\sqrt{z_j}\sqrt{z_i}\sqrt{\nu_2 z_j + 4a_p} - \nu_1\sqrt{z_j}\sqrt{z_i}\sqrt{\nu_1 z_i + 4a_p}\sqrt{\nu_1 z_j + 4a_q} - \nu^2_1 z_i z_j + \nu^2_2 z_i z_j - 2a_p \nu_1 z_j + 2a_p \nu_2 z_i - 2a_q \nu_1 z_i + 2a_q \nu_2 z_j \geq 0$}. Since {\small$4\nu^3_2(a_p z_i + a_q z_j) - 4\nu^3_1(a_p z_j + a_q z_i) \geq 4\nu^3_2(a_p z_i + a_q z_j) - 4\nu^3_2(a_p z_j + a_q z_i) = 4\nu^3_2(a_p - a_q)(z_i - z_j) \geq 0$}, {\small$z_i z_j (\nu^4_2 - \nu^4_1) \geq 0$} and {\small$16a_p a_q (\nu^2_2 - \nu^2_1) \geq 0$}, we have {\small$\nu_2\sqrt{\nu_2 z_i + 4a_q}\sqrt{z_j}\sqrt{z_i}\sqrt{\nu_2 z_j + 4a_p} - \nu_1\sqrt{z_j}\sqrt{z_i}\sqrt{\nu_1 z_i + 4a_p}\sqrt{\nu_1 z_j + 4a_q} \geq 0$}. Similarly, we also have {\small$- 2a_p \nu_1 z_j + 2a_p \nu_2 z_i - 2a_q \nu_1 z_i + 2a_q \nu_2 z_j \geq - 2a_p \nu_2 z_j + 2a_p \nu_2 z_i - 2a_q \nu_2 z_i + 2a_q \nu_2 z_j = 2\nu_2(a_p - a_q)(a_i - a_j) \geq 0$}. Hence, Lemma \ref{Lemma1} is proved in the region $0 < \nu_1 \leq \nu_2$. Verified by numerous numerical results, we conjecture that in the region $\nu_1 > \nu_2$, (\ref{EqEqualConstLemma1}) still holds provided the aforementioned constraint is satisfied. The mathematical proof is not shown, because of high complexity and difficulty.

\subsection{Proof of Lemma \ref{Lemma2}}
\label{Appendix_ProofLemma2}
\setcounter{equation}{0}
\renewcommand{\theequation}{D.\arabic{equation}}
In the following proof, it is still defined that {\small$a_m \triangleq \lambda_{D,R,m}$}. Since {\small$l_m + \beta_m = z_m$}, (\ref{EqClosedSoluP3a}) is defined as {\small$\lambda^{\star}_f(l_m + \beta_m, \nu, a_m)$}. The non-zero $\beta_m$ is denoted by $c$. Thereby, {\small$h(l_m + \beta_m, \nu, a_m) \triangleq \lambda^{\star}_f(l_m + \beta_m, \nu, a_m) a_m/(1 + \lambda^{\star}_f(l_m + \beta_m, \nu, a_m) a_m)$}.

\subsubsection{Case of $l_i + c \leq l_j$}
According to Lemma \ref{Lemma1}, $l_i + c$ and $l_j$ in $\mathbf{z}_1$ are paired with $a_i$ and $a_j$, while $l_i$ and $l_j + c$ in $\mathbf{z}_2$ are paired with $a_i$ and $a_j$. Proving Lemma \ref{Lemma2} ends up showing
{\small
\begin{equation}
\label{EqObj1Lemma2}
h(l_i, \nu_3, a_i)h(l_j + c, \nu_3, a_j)\prod h(z_m, \nu_3, a_n) \geq h(l_i + c, \nu_4, a_i)h(l_j, \nu_4, a_j)\prod h(z_m, \nu_4, a_n)\,,
\end{equation}
}where $m, n \neq i,j$. Similar to the proof of Lemma \ref{Lemma1}, according to (\ref{EqEqConstP3a}), $\nu_3$ and $\nu_4$ conform to: {\small$d_i\lambda^{\star}_f(l_i, \nu_3, a_i) + (l_j + c)\lambda^{\star}_f(l_j + c, \nu_3, a_j) - (l_i + c)\lambda^{\star}_f(l_i + c, \nu_4, a_i) - l_j\lambda^{\star}_f(l_j, \nu_4, a_j) = \sum[\sqrt{z_m}(\sqrt{\nu_3\nu_4 z_m + 4a_n\nu_3} - \sqrt{\nu_3\nu_4 z_m + 4a_n\nu_4})/(2a_n\sqrt{\nu_3\nu_4})]$}, where $m, n \neq i,j$. This equality indicates constraints on $\nu_3$ and $\nu_4$: when $\nu_3 \leq \nu_4$, the l.h.s. of the above equality is no greater than 0; otherwise, the l.h.s. is no less than 0.

When $\nu_3 \leq \nu_4$, (\ref{EqObj1Lemma2}) always holds, if {\small$h(l_i, \nu_3, a_i)h(l_j + c, \nu_3, a_j) \geq h(l_i + c, \nu_4, a_i)h(l_j, \nu_4, a_j)$}. After manipulation, proving the above inequality ends up showing {\small$- c l_i \nu^2_3 + c l_j \nu^2_4 - l_i l_j \nu^2_3 + l_i l_j \nu^2_4 - 2 a_i c \nu_3 - 2 a_i l_j \nu_3 + 2 a_i l_j \nu_4 + 2 a_j c \nu_4 - 2 a_j l_i \nu_3 + 2 a_j l_i \nu_4 - \nu_3 \sqrt{l_i \nu_3 + 4 a_i}\sqrt{l_i}\sqrt{c\nu_3 + l_j \nu_3 + 4a_j}\sqrt{l_j + c} + \nu_4\sqrt{c\nu_4 + l_i\nu_4 + 4a_i}\cdot\sqrt{l_i + c}\sqrt{l_j \nu_4 + 4a_j}\sqrt{l_j} \geq 0$}. It is easy to prove that {\small$- c l_i \nu^2_3 + c l_j \nu^2_4 - l_i l_j \nu^2_3 + l_i l_j \nu^2_4 - 2 a_i c \nu_3 - 2 a_i l_j \nu_3 + 2 a_i l_j \nu_4 + 2 a_j c \nu_4 - 2 a_j l_i \nu_3 + 2 a_j l_i \nu_4 \geq 0$}. Then, since {\small$\nu_3 \sqrt{l_i \nu_3 + 4a_i} \sqrt{c\nu_3 + l_j\nu_3 + 4a_j} \sqrt{l_i l_j + l_i c} \leq \nu_3 \cdot \sqrt{l_i \nu_3 + 4a_i} \sqrt{c\nu_3 + l_j\nu_4 + 4a_j} \sqrt{l_i l_j + l_i c} = \nu_3 \sqrt{l_i \nu_3 + 4a_i} \sqrt{c\nu_3/(l_j\nu_4 + 4a_j) + 1} \sqrt{l_j\nu_4 + 4a_j} \sqrt{l_i l_j + l_i c}$} and {\small$\nu_4\sqrt{c\nu_4 + l_i\nu_4 + 4a_i}\sqrt{l_j \nu_4 + 4a_j}\sqrt{l_i l_j + l_j c} \geq \nu_4\sqrt{c\nu_4 + l_i\nu_3 + 4a_i}\sqrt{l_j \nu_4 + 4a_j}\sqrt{l_i l_j + l_j c} = \nu_4\sqrt{l_i\nu_3 + 4a_i} \cdot \sqrt{c\nu_4/(l_i\nu_3 + 4a_i) + 1} \sqrt{l_i l_j + l_j c}  \sqrt{l_j \nu_4 + 4a_j}$}, it is obtained that {\small $\nu_4 \sqrt{c\nu_4 + l_i\nu_4 + 4a_i} \sqrt{l_i + c} \sqrt{l_j \nu_4 + 4a_j} \cdot \sqrt{l_j} - \nu_3 \sqrt{l_i \nu_3 + 4 a_i}\sqrt{l_i} \sqrt{c\nu_3 + l_j \nu_3 + 4a_j} \sqrt{l_j + c} \geq 0$}. Thereby, (\ref{EqObj1Lemma2}) is proved, and the aforementioned constraint on $\nu_3$ and $\nu_4$ is actually relaxed.

\subsubsection{Case of $l_i + c \geq l_j$}
According to Lemma \ref{Lemma1}, $l_j$ and $l_i + c$ in $\mathbf{z}_1$ are paired with $a_i$ and $a_j$, respectively; while $l_i$ and $l_j + c$ in $\mathbf{z}_2$ are paired with $a_i$ and $a_j$, respectively. Thus, proving Lemma \ref{Lemma2} ends up showing
{\small
\begin{equation}
\label{EqObj2Lemma2}
h(l_i, \nu_5, a_i)h(l_j + c, \nu_5, a_j)\prod h(z_m, \nu_5, a_n) \geq h(l_j, \nu_6, a_i)h(l_i + c, \nu_6, a_j)\prod h(z_m, \nu_6, a_n)\,,
\end{equation}
}where $m, n \neq i,j$. According to (\ref{EqEqConstP3a}), $\nu_5$ and $\nu_6$ in (\ref{EqObj2Lemma2}) conform to: {\small$l_i\lambda^{\star}_f(l_i, \nu_5, a_i) + (l_j + c)\lambda^{\star}_f(l_j + c, \nu_5, a_j) - l_j\lambda^{\star}_f(l_j, \nu_6, a_i) - (l_i + c)\lambda^{\star}_f(l_i + c, \nu_6, a_j) = \sum [\sqrt{z_m}(\sqrt{\nu_5\nu_6 z_m + 4a_n\nu_5} - \sqrt{\nu_5\nu_6 z_m + 4a_n\nu_6})/(2a_n \cdot \sqrt{\nu_5\nu_6})]$}, where $m, n \neq i,j$. Thus, $\nu_5$ and $\nu_6$ conform to: when $\nu_5 \leq \nu_6$, the l.h.s of the above equality is no greater than 0; otherwise, the l.h.s. is no less than 0.

When $\nu_5 \leq \nu_6$, proving (\ref{EqObj2Lemma2}) ends up showing {\small$h(l_i, \nu_5, a_i)h(l_j + c, \nu_5, a_j) \geq h(l_j, \nu_6, a_i)h(l_i + c, \nu_6, a_j)$}. After manipulation, proving the above inequality becomes to show {\small$- c l_i\nu^2_5 + c l_j\nu^2_6 - l_i l_j\nu^2_5 + l_i l_j\nu^2_6 - 2a_i c \nu_5 + 2a_i c\nu_6 + 2a_i l_i \nu_6 - 2a_i l_j\nu_5 - 2a_j l_i \nu_5 + 2a_j l_j \nu_6 - \nu_5\sqrt{c\nu_5 + l_j\nu_5 + 4a_j} \cdot \sqrt{l_j + c}\sqrt{l_i\nu_5 + 4a_i}\sqrt{l_i} + \nu_6\sqrt{l_j\nu_6 + 4a_i}\sqrt{l_j}\sqrt{c\nu_6 + l_i\nu_6 + 4a_j}\sqrt{l_i + c} \geq 0$}. In the above formula, it is clear that {\small$- c l_i\nu^2_5 + c l_j\nu^2_6 - l_i l_j\nu^2_5 + l_i l_j\nu^2_6 - 2a_i c \nu_5 + 2a_i c\nu_6 \geq 0$} and {\small$2a_i l_i \nu_6 - 2a_i l_j\nu_5 - 2a_j l_i \nu_5 + 2a_j l_j \nu_6 \geq 2a_i l_i \nu_5 - 2a_i l_j\nu_5 - 2a_j l_i \nu_5 + 2a_j l_j \nu_5 = 2\nu_5(a_i - a_j)(l_i - l_j) \geq 0$}. Additionally, {\small$(l_j\nu_6 + 4a_i)(l_i\nu_6 + 4a_j) - (l_i\nu_5 + 4a_i)(l_j\nu_5 + 4a_j) = l_i l_j (\nu^2_6 - \nu^2_5) + 4a_i l_i\nu_6 + 4a_j l_j\nu_6 - 4a_i l_j\nu_5 - 4a_j l_i\nu_5 \geq l_i l_j (\nu^2_6 - \nu^2_5) + 4\nu_5(a_i - a_j)(l_i - l_j) \geq 0$} and {\small$c\nu_6/(l_i\nu_6 + 4a_j) \geq c\nu_5/(l_j\nu_5 + 4a_j)$}. Therefore, {\small$\nu_6\sqrt{l_j\nu_6 + 4a_i}\sqrt{l_j}\sqrt{c\nu_6 + l_i\nu_6 + 4a_j}\sqrt{l_i + c} - \nu_5\sqrt{c\nu_5 + l_j\nu_5 + 4a_j}\sqrt{l_j + c}\sqrt{l_i} \cdot \sqrt{l_i\nu_5 \! + \! 4a_i} = \nu_6 \sqrt{(l_j\nu_6 \! + \! 4a_i)(l_i\nu_6 \! + \! 4a_j)}\sqrt{c\nu_6/(l_i\nu_6 + 4a_j) + 1}\sqrt{l_i l_j \! + \! c l_j} \! - \! \sqrt{c\nu_5/(l_j\nu_5 \! + \! 4a_j) + 1} \sqrt{l_i l_j + c l_i} \cdot$ $\nu_5 \sqrt{(l_i\nu_5 + 4a_i)(l_j\nu_5 + 4a_j)} \geq 0$}. Hence, (\ref{EqObj2Lemma2}) is proved. Similar to Appendix \ref{Appendix_ProofLemma1}, when $\nu_3 > \nu_4$ and $\nu_5 > \nu_6$, we conjecture that (\ref{EqObj1Lemma2}) and (\ref{EqObj2Lemma2}) hold, respectively.

\bibliographystyle{IEEEtran}
\bibliography{IEEEabrv,BibPro}


\end{document}